\documentclass[12pt]{article}
\usepackage{amsmath,amssymb,amsfonts,amsthm,bbm}
\usepackage{epic,eepic,epsfig,longtable}
\usepackage{graphicx,psfrag,epsf}
\usepackage{multirow,verbatim}
\usepackage{floatrow}
\usepackage{subcaption}
\usepackage{paralist}
\usepackage{latexsym}
\usepackage{comment}
\usepackage{array}
\usepackage{enumerate}
\usepackage{setspace}
\usepackage{enumitem}
\usepackage{CJK}
\usepackage{color}
\usepackage{booktabs}
\usepackage{soul}
\usepackage{rotating}
\usepackage{mathtools}
\usepackage[authoryear, round]{natbib}

\def\b#1{\textcolor{blue}{#1}}

\usepackage{url} 
\newcommand{\blind}{0}
\providecommand{\abs}[1]{\left\lvert#1\right\rvert}
\providecommand{\norm}[1]{\left\lVert#1\right\rVert}

\numberwithin{equation}{section}

\renewcommand{\hat}{\widehat}

\renewcommand{\hat}{\widehat}

\newcommand{\bfm}[1]{\ensuremath{\mathbf{#1}}}

   \def\bA{\bfm A}  
   \def\bB{\bfm B}

\def\be{\bfm e}     
    \def\FF{\mathbb{F}}
\def\bg{\bfm g}   \def\bG{\bfm G}  
\def\bh{\bfm h}     
   \def\bI{\bfm I}  
   \def\bJ{\bfm J}

\def\bm{\bfm m}   \def\bM{\bfm M}

   \def\bP{\bfm P}  
\def\bq{\bfm q}     
\def\br{\bfm r}

\def\bu{\bfm u}   \def\bU{\bfm U}  
     
\def\bw{\bfm w}   \def\bW{\bfm W}  
   \def\bX{\bfm X}  
   \def\bY{\bfm Y}  
   \def\bZ{\bfm Z}  

\def\b1{\bfm 1}

\newcommand{\bfsym}[1]{\ensuremath{\boldsymbol{#1}}}

 \def\bbeta{\bfsym \beta}
              \def\bGamma{\bfsym \Gamma}

 \def\bmu{\bfsym {\mu}}                 
 
 \def\btheta{\bfsym {\theta}}           
           \def\bepsilon{\bfsym \varepsilon}
 \def\bsigma{\bfsym \sigma}             \def\bSigma{\bfsym \Sigma}
         \def\bLambda {\bfsym {\Lambda}}
           \def\bOmega {\bfsym {\Omega}}

\def\1{\bfsym{1}}

\DeclareMathOperator{\argmin}{argmin}

\DeclareMathOperator{\diag}{diag}

\DeclareMathOperator*{\argminU}{argmin}

\def\newpage{\vfill\eject}

\def\today{\ifcase\month\or
  January\or February\or March\or April\or May\or June\or
  July\or August\or September\or October\or November\or December\fi
  \space\number\day, \number\year}

\newdimen\biblioindent    \biblioindent=30pt

 at 8truept

\newcommand{\beq}{\begin{equation}}
  \newcommand{\eeq}{\end{equation}}
\newcommand{\beqn}{\begin{eqnarray}}
  \newcommand{\eeqn}{\end{eqnarray}}
\newcommand{\beqnn}{\begin{eqnarray*}}
  \newcommand{\eeqnn}{\end{eqnarray*}}

\def\tilde{\widetilde}

\def\FF{\mathcal{F}}
\def\[{\left [}  \def\]{\right ]} \def\({\left (}  \def\){\right )}
 \def\endpf{$\blacksquare$}
\def\hat{\widehat}
\makeatletter
\def\underbar#1{\underline{\sbox\tw@{$#1$}\dp\tw@\z@\box\tw@}}
\makeatother

\newtheorem{assumption}{Assumption}
\newtheorem{theorem}{Theorem}

\newtheorem{proposition}{Proposition}
\theoremstyle{definition}

\newtheorem{remark}{Remark}

\def \diag {\mathrm{diag}} 
\def \det {\mathrm{det}}

\def\spacingset#1{\renewcommand{\baselinestretch}%
{#1}\small\normalsize} \spacingset{1}

\begin{document}

\def\spacingset#1{\renewcommand{\baselinestretch}%
{#1}\small\normalsize} \spacingset{1}

\if0\blind
{
  \title{\bf Dynamic Realized Minimum Variance Portfolio Models}
  \author{Donggyu Kim \\
    College of Business,\\ Korea Advanced Institute of Science and Technology (KAIST)\\
    and \\
    Minseog Oh \\
    College of Business,\\ Korea Advanced Institute of Science and Technology (KAIST)}
  \maketitle
} \fi

\if1\blind
{
  \bigskip
  \bigskip
  \bigskip
  \begin{center}
    {\LARGE\bf Title}
\end{center}
  \medskip
} \fi

\bigskip
\begin{abstract}
  This paper introduces a dynamic minimum variance portfolio (MVP) model using nonlinear volatility dynamic models, based on high-frequency financial data.
  Specifically, we impose an autoregressive dynamic structure on MVP processes, which helps  capture the MVP dynamics directly.
  To evaluate the dynamic MVP model, we estimate the inverse volatility matrix using the constrained $\ell_1$-minimization for inverse matrix estimation (CLIME) and calculate daily realized non-normalized MVP weights.
  Based on the realized non-normalized MVP weight estimator, we propose the dynamic MVP model, which we call the dynamic realized minimum variance portfolio (DR-MVP) model.
  To estimate a large number of parameters, we employ the least absolute shrinkage and selection operator (LASSO) and predict the future MVP and establish its asymptotic properties.
  Using high-frequency trading data, we apply the proposed method to MVP prediction.
\end{abstract}

\noindent%
\vfill

\newpage
\spacingset{1.8} 
\section{Introduction} \label{SEC-1}

The minimum variance portfolio (MVP) has received growing attention \citep{demiguel2009generalized, fan2012vast, ledoit2017nonlinear}.
Unlike the mean-variance portfolio proposed in \citet{markowitz1952portfolio}, the MVP  avoids the difficulty of estimating expected returns \citep{merton1980estimating} and only requires accurate covariance matrix estimators.
Furthermore, empirical studies have found that the MVP can enjoy both lower risk and higher return, compared with some benchmark portfolios \citep{chan1999portfolio, clarke2006minimum, haugen1991efficient, jagannathan2003risk}.
Several well-performing non-parametric realized volatility estimators have been developed since high-frequency financial data became available.
Examples of the finite number of assets include  
two-time scale realized volatility \citep{zhang2005tale}, 
multi-scale realized volatility \citep{zhang2006efficient, zhang2011estimating}, 
wavelet estimator \citep{fan2007multi}, 
kernel realized volatility \citep{barndorff2008designing, barndorff2011multivariate},
pre-averaging realized volatility \citep{christensen2010pre, jacod2009microstructure}, 
quasi-maximum likelihood estimator \citep{ait2010high, xiu2010quasi}, 
local method of moments \citep{bibinger2014estimating}, 
and robust pre-averaging realized volatility \citep{fan2018robust, shin2023adaptive}.
On the other hand, estimation methods for large volatility matrices have been developed based on the sparse volatility matrix structure and approximate factor models \citep{ait2017using, dai2019knowing, fan2012vast, kim2018adaptive, kim2018large,  pelger2019large, tao2013optimal}.
Using these non-parametric volatility matrix estimation procedures, we can accurately estimate large volatility matrices within a relatively short time period, such as one day or a few days. 
This makes it possible to  estimate past optimal MVP weights.
For example, \citet{fan2012vast} studied the MVP under the gross-exposure constraint, and  \citet{cai2020high} proposed the estimation of high-dimensional MVP using the   constrained $\ell_1$-minimization for inverse matrix estimation (CLIME) \citep{cai2011constrained}.

\begin{figure}[!ht] 
	\centering
	\includegraphics[width = 0.9\textwidth]{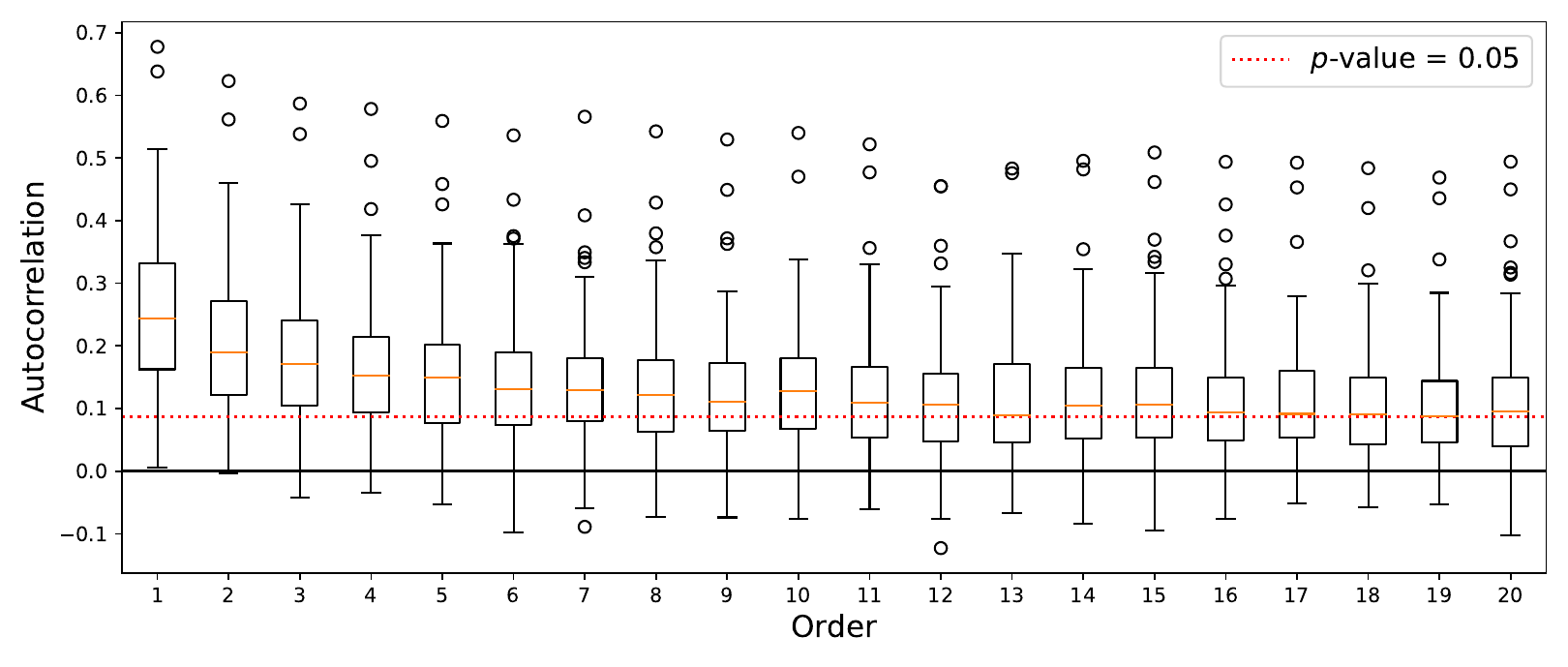} 
	\caption{Box-plots of the first- to twentieth-order autocorrelations for each non-normalized MVP weight.}
	\label{ACF}
\end{figure}

In practice, we often need to predict future MVPs given current information.
To predict the MVP, we often first predict the large volatility matrix and then we solve the minimum variance portfolio problem with the predicted large volatility matrix estimator \citep{kim2019factor, shin2021factor}. 
However, this approach does not reflect the MVP dynamics directly, which may cause some loss of explanatory power from them. 
To check the MVP dynamics, we draw the box-plots for the autocorrelations for each non-normalized MVP weight in Figure \ref{ACF}, where the non-parametric MVP weight estimation method is presented in Section \ref{sec-MVP}. 
The horizontal red dotted line in Figure \ref{ACF} indicates the significant autocorrelation level with 5\% significance.
From Figure \ref{ACF}, we find that the weight has autoregressive dynamics.
Thus, by directly modeling these dynamics, we may be able to better explain the MVP dynamics.
This fact leads us to develop an MVP dynamic model. 

In this paper, we propose a dynamic MVP model based on high-frequency financial data. 
Specifically, we first investigate a non-parametric MVP estimation procedure.
To tackle the curse of dimensionality, we impose a sparse structure on the inverse matrices of daily integrated volatility matrices. 
Using high-frequency financial data, we estimate daily integrated volatility matrices and then, to accommodate the sparsity, employ the CLIME procedure \citep{cai2011constrained}. 
With the inverse matrix estimator, we estimate non-normalized MVP weights and establish their asymptotic properties.
Then, to capture the autoregressive dynamic structure of these weights, we propose a dynamic MVP  model. 
Since the proposed model explains low-frequency dynamics based on high-frequency financial data,
we connect the low-frequency dynamics with a continuous diffusion process.
That is, the proposed model has the rigorous mathematical background when it comes to incorporating high-frequency financial data to analyze low-frequency dynamics. 
We call this the dynamic realized minimum variance portfolio (DR-MVP) model.
To evaluate the proposed DR-MVP model, we need to estimate a large number of parameters that come from a large number of assets and long period time series. 
To handle this curse of dimensionality, we assume that the model parameters are sparse and employ an $\ell_{1}$-penalty, such as the least absolute shrinkage and selection operator (LASSO) \citep{tibshirani1996regression}. 
Based on the predicted non-normalized MVP weights, we estimate the conditional expected future MVP and show its consistency. 
In the empirical study, we find the benefit of adopting the DR-MVP model.

The rest of the paper is organized as follows.
Section \ref{SEC-2}  introduces the  DR-MVP model based on high-dimensional  It\^{o} diffusion processes and investigates its properties.
Section \ref{SEC-3} proposes the non-parametric MVP estimation procedure and the parameter estimation method for the high-dimensional DR-MVP model, and establishes its asymptotic properties.
In Section \ref{SEC-4}, we conduct a simulation study to check the finite sample performance of the proposed estimator.
In Section \ref{SEC-5}, we apply the estimation method to high-frequency trading data.
The conclusion is presented in Section \ref{SEC-6}, and we collect the proofs in the online Appendix.

\section{Realized Minimum Variance Portfolio} \label{SEC-2}

Let $\bX_t= (X_{1,t} , \ldots, X_{p,t})^\top$ be the vector of true log-prices of $p$ assets at time $t$.
The log-price follows  the following continuous diffusion process:
\begin{equation}\label{model-1}
  d \bX_t= \bmu_t dt + \bsigma_t d {\mathcal{B}}_t,
\end{equation}
where $\bmu_t$ and $\bsigma_t$ are drift and instantaneous volatility processes, and ${\mathcal{B}}_t$ is a $p$-dimensional standard Brownian motion. 
Stochastic processes $\bX_t$, $\bmu_t$,  and $\bsigma_t$ are defined on a filtered probability space $(\Omega, \FF, \{\FF_t, t \in [0, \infty)\},  P)$ with filtration $\FF_t$ satisfying the usual conditions, such as c\`adl\`ag and  adapted processes.
We denote $\bSigma_{t} = \bsigma_{t}\bsigma_{t}^{\top}$; that is, $\bsigma_{t}$ is square root of positive semidefinite spot covariance matrix $\bSigma_{t}$.
  
We construct a portfolio at the beginning of the market and hold it during trading hours. 
Let $\bar{\bw}_d  $ be the realized optimal MVP at day $d$, which satisfies 
\begin{equation} \label{minimum-port}
	\bar{\bw}_d = \arg \min_{\bh} \bh^{\top} \bGamma_d \bh \quad \text{ s.t. } \b1 ^{\top} \bh =1, 
\end{equation}
where $\b1 = (1,\ldots,1 )^{\top} \in \mathbb{R}^{p}$, and $\bGamma_d = \int_{d-1}^d \bsigma_t \bsigma_t^{\top} dt$ is the daily integrated volatility.
Then, we have the following closed form:
\begin{equation*}
  \bar{\bw}_d = \frac{1}{\b1 ^{\top} \bGamma_d ^{-1}  \b1 } \bw_d, \quad \text{where} \quad \bw_d =  \bGamma_d ^{-1} \b1. 
\end{equation*}
The MVP weight vector $\bar{\bw}_d$ minimizes the quadratic variation of the portfolio.
Specifically, we have the portfolio $\bar{\bw}_d ^{\top}   \bX_{t}$ during $t \in [d-1, d)$, and its quadratic variation for the $d$th day is  
\begin{equation*}
 	\[  \bar{\bw}_d ^{\top}   \bX_{t} \]_d -\[  \bar{\bw}_d ^{\top}   \bX_{t} \]_{d-1}   =  \bar{\bw}_d ^{\top} \bGamma_d \bar{\bw}_d  \quad \text{ a.s.}
\end{equation*}
Since $\bar{\bw}_d$ satisfies \eqref{minimum-port}, the quadratic variation obtains the minimum value among the class of the weight $\bh$ such that $\b1 ^{\top} \bh=1$.
We call $\bar{\bw}_d$ the realized MVP.

In this paper, we investigate the  realized MVP dynamics and construct dynamic models to accommodate them. 
We first investigate the non-normalized weight $\bw_d =  \bGamma_d ^{-1} \b1$. 
We consider the following inverse matrix dynamic model:
\begin{equation}\label{eq:inversematrixmodel}
	\bOmega_{d}= \bG_{d}(\btheta) + \bU_d,
\end{equation}
where $\bG_{d}(\btheta)$ is $\FF_{d-1}$-adapted, $\btheta$ is a model parameter, and $E \[ \bU_{d} \middle| \FF_{d-1}  \] =0  $ a.s.
Then, we have the following dynamic realized portfolio:
\begin{equation}\label{eq:g-G}
  \bw_d =  \bg_{d} (\btheta) + \bepsilon_{d}, 
\end{equation}
where  $\bg_{d} (\btheta)= \bG_{d} (\btheta) \b1$ , $\bepsilon_d= \bU_d \b1$, and $\bepsilon_{d}$'s are martingale differences.
Thus, the conditional expected value of $\bw_d$ is $E \[ \bw_d \middle | \FF_{d-1} \] = \bg_{d} (\btheta)$ a.s.
We call this the dynamic realized minimum variance portfolio (DR-MVP) model.
We note that $\bg_{d} (\btheta)$  can be any function of $\FF_{d-1}$-adapted variables, and one of the possible choices is a form of autoregressive (AR), as follows: 
\begin{equation}\label{eq:g-AR}
	\bg_{d} (\btheta) = \bbeta_0 + \sum_{j=1}^{q} \bbeta_j \bw_{d-j},
\end{equation}
where $\bbeta_{0} \in \mathbb{R}^{p}$ and $\bbeta_{j}$'s are $p$-by-$p$ matrices.
There are many structures of $\bG_d(\btheta)$ satisfying \eqref{eq:g-G}. 
For example, $\bG_{d}(\btheta)$ can be the form of BEKK$(0, q)$ \citep{engle1995multivariate} that can ensure its positive semidefiniteness.
Specifically, \eqref{eq:g-AR} holds if 
\begin{equation}\label{eq:G-AR}
  \bG_{d}(\btheta) = \bB_{0}\bB_{0}^{\top} + \sum_{j=1}^{q} \bB_{j} \bOmega_{d-j} \bB_{j}^{\top} 
  ,
\end{equation}
where $[\bB_{j}]_{kl} = \bbeta_{j,kl} / \sqrt{\sum_{r=1}^{p} \bbeta_{j,rl}}$ and $\bbeta_0 = \bB_0 \bB_0^{\top} \b1$ under some  conditions on $\bbeta$ or $\bOmega$.
The toy examples are as follows.
One of the sufficient conditions is that $\bOmega$ is a diagonal matrix.
The diagonal condition allows $\bbeta$ to be of any form, while the correlation between returns of assets should be zero, which is too restrictive.
Another sufficient condition is that all column-sums of $\bbeta$ are the same. 
For example, $\bbeta^{\top}$ has an eigenvector of $p^{-1/2}\b1$.
We note that if the corresponding eigenvalue is bounded, then its effect on the inverse matrix is relatively small.
Under the $\bbeta$ restrictive condition, we do not need a condition for the inverse matrix  $\bOmega$. 
Alternatively, to relieve the condition, we can consider a compromise condition that balances the diagonal condition and $\bbeta$ restrictive condition. 
Specifically, the compromise condition is that if $\sum_{t=1}^{p} \bbeta_{tk} \neq \sum_{t=1}^{p} \bbeta_{tl}$, then $\bOmega_{kl}$ is zero.
Under this condition, \eqref{eq:g-AR} holds if \eqref{eq:G-AR} is satisfied.
These toy examples illustrate that there exists an inverse matrix dynamic model satisfying the DR-MVP model.
In general, we can associate the inverse matrix dynamics model of the BEKK form with the DR-MVP model by dividing the integrated volatility matrix into two parts.
One of the matrices follows an inverse matrix dynamic model with $\bG_{d}(\btheta)$ satisfying \eqref{eq:G-AR}, and the other matrix serves as the supplementary component to make the MVP weights satisfy \eqref{eq:g-AR}.
We provide the specific condition in the online Appendix \ref{cond-general} and utilize it in the simulation study.
 
We construct a diffusion volatility matrix process that satisfies the inverse matrix dynamic model.
In the following theorem, we show the existence of an instantaneous volatility process satisfying \eqref{eq:inversematrixmodel}. 

\begin{theorem}\label{prop:DRMVPvol}
  Suppose that the volatility matrix process satisfies, for any $d \in \mathbb{N}$ and $t \in (d-1, d]$, 
  \begin{equation}\label{eq:DRMVPvol}
    \bSigma_{t}  = 2 \bA_{t-} - \bA_{t-} \bSigma_{d-1}^{-1} \bA_{t-} + (4 \langle t \rangle  - 3 \langle t \rangle ^{2}) \bA_{t-} (\bSigma_{d-1}^{-1} - \bG_{d}(\btheta)) \bA_{t-} - \bA_{t-} \bM_{t} \bA_{t-} ,
  \end{equation}
  where
  $\langle t \rangle = t - d + 1$, $\bA_{t} =   \langle t \rangle ^{-1} \left(  \int_{d-1}^{t} \bSigma_{s} ds \right)$, $\bM_{t} = \int_{d-1}^{t} d\bZ_{t}$,  and $\bZ_{t}$ is a symmetric matrix-valued martingale process.
  Then, its inverse integrated volatility matrices follow the inverse matrix dynamic model \eqref{eq:inversematrixmodel};
  that is, $\bOmega_{d} = (\int_{d-1}^{d} \bSigma_{t} dt)^{-1} = \bG_{d} (\btheta) + \bU_{d}$, where $\bU_{d} = \int_{d-1}^{d} \bM_{t}  dt$ is a martingale difference.
\end{theorem}

Theorem \ref{prop:DRMVPvol} indicates that as long as an instantaneous volatility process satisfies \eqref{eq:DRMVPvol}, its inverse integrated volatility matrix over the $d$th period can be decomposed into the conditional expectation $\bG_{d}(\btheta)$ and the martingale difference $\bU_{d}$.
Meanwhile, the proposed instantaneous volatility process should satisfy the positive semidefiniteness for any time $t$. 
In the online Appendix \ref{cond-pd}, we provide sufficient conditions that ensure the positive semidefiniteness of the proposed volatility process.
On the other hand, in terms of volatility processes, the proposed instantaneous volatility model accommodates the nonlinear dynamics, which helps capture the MVP dynamics. 
This is the main difference from the existing parametric high-frequency-based large volatility dynamic models \citep{kim2019factor, shin2021factor}.
The empirical study supports the nonlinear volatility model (see Section \ref{SEC-5}).

The goal of this paper is to predict the future MVP given the current available information.
Thus, the variable of interest is the conditional expected portfolio, namely 
\begin{equation*}
	E \[ \bar{\bw}_d \middle | \FF_{d-1} \] =  E\[ \frac{\bw_{d} }{\b1^{\top} \bw_d } \middle | \FF_{d-1}\].
\end{equation*}
Due to the random fluctuation term, this is not exactly the same as the normalized $\bg_{d} (\btheta)$, $\frac{\bg_{d} (\btheta)}{\b1^{\top} \bg_{d} (\btheta)}$.
We investigate the error coming from this approximation.
To do this, we need the following technical conditions.

\begin{assumption}\label{assumption1} ~
\begin{enumerate}
\item [(a)]  For any $d$, there exist positive constants $C_1$ and $C_2$ such that $C_1 \leq p^{-1} \b1 ^{\top} \bG_d(\theta) \b1 \leq  C_2$  and $ C_1  \leq  p^{-1}  \b1 ^{\top} \bOmega_d \b1 \leq C_2$, where $\bG_d(\btheta) = E \[ \bOmega_{d} \middle| \FF_{d-1} \]$.

\item [(b)]  $ \epsilon_{i, d} $'s are sub-Gaussian, and we have $E \[ ( \b1^{\top} \bepsilon_{d} )^2    \middle | \FF_{d-1} \] \leq  p C$ almost surely, where $\bepsilon_{d} = (\epsilon_{1,d},\ldots,  \epsilon_{p,d})^{\top}$. 
\end{enumerate}
\end{assumption}

\begin{remark}
Assumption \ref{assumption1}(a) is related to the eigenvalue condition for the inverse matrix. 
That is, if eigenvalues are bounded and strictly bigger than zero, Assumption \ref{assumption1}(a) is satisfied. 
However, when considering factor models, we can have some eigenvalues that converge to zero with $p^{-1}$ order. 
Furthermore, if $ p^{-1/2}  \b1 ^{\top}$ is close to the factor loading matrix space, Assumption \ref{assumption1}(a) may not hold. 
However, in this case, we have that $ \b1 ^{\top} \bOmega_d \b1$ has the constant order, and $ \b1^{\top} \bepsilon_{d}$ also has the constant order.
Thus, we can obtain the same result in Proposition \ref{prop1}.
For simplicity, we impose Assumption \ref{assumption1}, and, in this paper, we implicitly assume that the non-normalized weight $\bw_d$ is a usual finite random vector that does not depend on $p$. 
If it does not hold, by multiplying the corresponding order, we can make the vector have a constant order.
Therefore, this condition is not strong. 
On the other hand, for the inverse matrix, we assume the sparsity (see  \eqref{sparsity}), which implies the sparsity of the random fluctuation part $\bU_d$.
Thus, it is not restrictive to assume that the row-wise sum of $\bU_d$ has a finite random variable (Assumption  \ref{assumption1}(b)).
In fact, if the random fluctuation elements $U_{ij d}$'s are cross-sectionally independent or weakly dependent, we can obtain Assumption  \ref{assumption1}(b). 
\end{remark} 
The following proposition shows the error rate of the approximation.
\begin{proposition} \label{prop1}
Under Assumption \ref{assumption1}, we have
\begin{eqnarray}
	&&\left \|  E \[ \bar{\bw}_d \middle | \FF_{d-1}  \] - 	\frac{\bg_{d} (\btheta)}{\b1^{\top} \bg_{d} (\btheta)}  \right \| _{\max} = O_p \(  \left  \|  	\frac{\bg_{d} (\btheta)}{\b1^{\top} \bg_{d} (\btheta)}  \right \|_{\max} p^{-1/2} + p^{-3/2} \sqrt{\log p}     \) , \label{prop1-result1} \\
	&&\left  \|  E \[ \bar{\bw}_d \middle | \FF_{d-1} \] - 	\frac{\bg_{d} (\btheta)}{\b1^{\top} \bg_{d} (\btheta)}  \right \| _{1} = O_p \(   \left \|  	\frac{\bg_{d} (\btheta)}{\b1^{\top} \bg_{d} (\btheta)}  \right  \|_{1}    p^{-1/2}   \).\label{prop1-result2}
\end{eqnarray}
\end{proposition}

Proposition \ref{prop1} indicates that the conditional expected MVP can be estimated well by the normalized $\bg_{d} (\theta)$ as the number of assets goes to infinity.
The error is $p^{-1/2}$ times smaller than the normalized $\bg_{d} (\theta)$, except for the case that the rate of the $\norm{\bg_{d} (\btheta)}_{\max}$ is less than $\sqrt{\log p}$. 
Thus, we use $ \frac{\bg_{d} (\btheta)}{\b1^{\top} \bg_{d} (\btheta)}$ as the conditional expected value of the realized MVP. 
In the following section, we discuss how to estimate $\bg_{d-1} (\btheta)$.

\section{Estimation Procedure} \label{SEC-3}
The intraday log-prices for the $d$th day are observed at $t_{d,i}, i=1,\ldots, m_d$, where $d-1=t_{d,0} < t_{d,1} < \cdots < t_{d, m_d} = d$.  
Unfortunately, due to market microstructure noises, true high-frequency observations,  $\bX_{t_{d,i}}$'s, are not observed. 
To account for the market microstructure noises, we assume that the observed log-prices $\bY_{t_{d,i}}$ have the following additive noise structure:
\begin{equation*}
 	\bY_{t_{d,i}}= \bX_{t_{d,i}} + \be_{t_{d,i}}, \quad \text{for } d=1,\ldots, N, i=1,\ldots, m_d, 
\end{equation*}
where $\bX_t$ is the true log-price and $\be_{t_{d,i}}$'s are microstructure noises with mean zero.

To remove the effect of market microstructure noises,  researchers have constructed nonparametric realized volatility estimators that take advantage of subsampling and local averaging techniques so that the integrated volatility can be estimated consistently and efficiently \citep{ait2010high, christensen2010pre, fan2018robust, jacod2009microstructure, xiu2010quasi,   zhang2006efficient, zhang2011estimating}. 
They demonstrated that the realized volatility estimator  has the convergence rate of $m^{-1/4}$, which is known as the optimal convergence rate with the presence of the microstructure noise. 
In this paper, we employ the  pre-averaging realized volatility estimator \citep{christensen2010pre, jacod2009microstructure} in the numerical study. 
 We denote the realized volatility matrix estimator by $\hat{\bGamma}$.

 \subsection{Non-parametric Realized Minimum Variance Portfolio Estimator} \label{sec-MVP}
 
 We first introduce some notations.
For any given  $p_1$ by  $p_2$ matrix $\bM = \left(M_{ij}\right)$,
 the Frobenius norm of $\bM$ is denoted by $\|\bM\|_F = \sqrt{ \mathrm{tr}(\bM^{\top} \bM) }$, 
the matrix spectral norm $\|\bM\|_2$ is the square root of the largest eigenvalue of $\bM\bM^\top$, and  let
 \begin{equation*}
  	\|\bM\|_1 = \max\limits_{1 \leq j \leq p_2}\sum\limits_{i = 1}^{p_1}|M_{ij}|,\hspace{0.5cm} \|\bM\|_\infty = \max\limits_{1 \leq i \leq p_1}\sum\limits_{j = 1}^{p_2}|M_{ij}| , \hspace{0.5cm}
    \| \bM \| _{\max} = \max_{i,j} | M_{ij}|.
 \end{equation*}
$C$'s denote generic positive constants whose values are free of other parameters and may change from appearance to appearance.
 
 In this section, we introduce a  realized MVP estimation procedure. 
 To estimate the realized MVP, we first estimate the inverse matrix of the integrated volatility matrix. 
 We assume that the inverse matrix of the integrated volatility matrix, $\bGamma_d^{-1} = \bOmega_d = (\Omega_{ij, d})_{i,j=1,\ldots,p}$, satisfies the following sparsity condition:
  \begin{equation} \label{sparsity}
  	 \sup_{  d  } \max_{ 1 \leq i \leq p} \sum_{j=1}^p |\Omega_{ij, d} | ^{\delta} \leq s_{ p } \,  \text{ a.s.},
 \end{equation}
 where $\delta \in [0, 1)$  and  $s_{p}$ is diverging slowly with respect to $p$, such as $\log p$. 
 To accommodate this sparsity, we use the constrained $\ell_1$-minimization for inverse matrix estimation (CLIME) \citep{cai2011constrained} with the realized volatility matrix estimator.
 Specifically, let $\hat{\bOmega}_{d}$ be the solution for the following optimization problem:
\begin{equation} \label{CLIME}
	\min \| \bA\|_1 \quad \text{s.t.} \quad  \|    \hat{\bGamma}_d   \bA - \bI  \|_{\max} \leq \tau_m, 
\end{equation}
where $\tau_m$ is the tuning parameter  specified in Proposition \ref{thm1}. 
 Then, we estimate the realized non-normalized MVP estimator as $\hat{\bw}_d= \hat{\bOmega}_d \b1$.
To investigate its asymptotic behavior, we need the following technical conditions.
 
 \begin{assumption}\label{assumption2} ~
 \begin{enumerate}
 \item [(a)] There exists a large $C_a$ depending on given constant $a$ such that
 \begin{equation*}
 	P \( |\hat{\Gamma}_{ij, d} - \Gamma_{ij,d} | \geq C_a \sqrt{\log (p \vee N) } m^{-1/4}   \) \leq C (p \vee N )^a. 
 \end{equation*}

 \item [(b)] $ \max_{d \leq N} \| \bOmega_d\|_1 \leq C$ a.s. 
 \end{enumerate}
 \end{assumption}
 \begin{remark}
 	The sub-exponential tail condition in Assumption \ref{assumption2}(a) is often imposed to handle high-dimensional statistics. 
 	Under the bounded instantaneous and drift condition, we can obtain the sub-exponential condition \citep{Kim2016SPCA, tao2013optimal}.
 	In contrast, for heavy-tailed observations, by employing the truncation method, we can obtain Assumption \ref{assumption2}(a) \citep{fan2018robust, shin2023adaptive}.
 	Thus, Assumption \ref{assumption2}(a) is not restrictive. 
\end{remark}
 
 The following proposition establishes the convergence rates for the inverse matrix estimator and realized non-normalized MVP estimator.

\begin{proposition}\label{thm1}
  Under Assumption \ref{assumption2} and the sparsity condition \eqref{sparsity}, we choose $\tau_m = C_{\tau}  m^{-1/4}  \sqrt{\log (p \vee N) }$ for some large constant $C_{\tau}$.
  Then, we have 
  \begin{equation} \label{thm-result1}
 	  \max_{d \leq N} \| \hat{\bOmega}_d - \bOmega_d \| _{\max} \leq  C  \tau_m,   \quad   \quad   \max_{d \leq N} \| \hat{\bOmega}_d - \bOmega_d  \| _{1} \leq C s_p \tau_m ^{1-\delta}, \quad \text{and}
  \end{equation}
  \begin{equation}\label{thm-result2}
 		\max_{d \leq N} \| \hat{\bw}_d -\bw_{d} \| _{\max}  \leq C s_p \tau_m ^{1-\delta}
  \end{equation}
  with probability greater than $1 -(p\vee N)^{-c}$ for any given positive constant $c$.
\end{proposition}

\begin{remark}
Proposition \ref{thm1} indicates that the CLIME estimator has the convergence rate $m^{-1/4}   \sqrt{\log (p \vee N) }$. 
When microstructure noise is present, it is known that  $m^{-1/4}$ is the optimal rate \citep{tao2013optimal}. 
Thus, the CLIME estimator can obtain the optimal rate up to log order. 
On the other hand, since the realized non-normalized MVP estimator $\hat{\bw}_d$ is the form of the row-wise sum of the CLIME estimator, the convergence rate of  $\hat{\bw}_d$  is bounded by the $\ell_1$ norm bound of the CLIME estimator. 
When we consider the exact sparsity ($\delta=0$),  $\hat{\bw}_d$ has the convergence rate $s_p m^{-1/4}   \sqrt{\log (p \vee N) }$. 
The term $s_p$ is the number of non-zero elements, which is usually assumed to be negligible in high-dimensional statistics, such as $\log p$. 
Thus,  $\hat{\bw}_d$ has the optimal rate $m^{-1/4}$ up to log order. 
\end{remark}
 
\subsection{Model Parameter Estimator}
  
In this paper, we assume that  $\bg_d (\btheta)$ is a function of the past realized non-normalized MVP weights $\bw_{d-1} ,\ldots, \bw_{1}$.
That is, we consider a time series structure. 
We note that it would be straightforward to extend this to a more general model by including exogenous variables. 
We estimate  $\bg_d (\btheta)$ by using the plug-in method. 
Specifically, we use the realized non-normalized MVP estimator  $\hat{\bw}_{d}$'s instead of  $\bw_{d}$'s.
We denote this estimator by  $\hat{\bg}_d (\btheta)$.
For example, when we consider the AR$(q)$ structure,   we have
\begin{equation*}
	\hat{\bg}_{d} (\btheta) = (\hat{g}_{i,d} (\btheta_1), \ldots, \hat{g}_{p,d} (\btheta_p) ) ^{\top} = \bbeta_{0} +  \sum_{j=1}^{q} \bbeta_{j}  \hat{\bw}_{d-j}.
\end{equation*}
 In this section, since the empirical study supports the AR structure (see Figure \ref{ACF}), we derive asymptotic theorems based on the AR$(q)$ model for simplicity.  
However, for any well-defined $\bg_d(\theta)$, we can derive similar results under some regularity conditions.

 The model is based on the high-dimensional vector autoregressive (VAR) model.
 Thus, we suffer from the curse of dimensionality. 
 To tackle this obstacle, we assume that the coefficient $\bbeta_j$ is sparse (see Assumption \ref{assumption3}(b)).
 To accommodate the sparsity, we employ the LASSO  for each weight as follows: 
\begin{equation*}
  \hat{\btheta_i}= \arg \min_{\btheta_i}   \mathcal{L}_{n,i} (\btheta_i)  + \lambda_n  \|\btheta_i\|_1 ,
\end{equation*}
where  $\mathcal{L}_{n,i} (\btheta_i) = \frac{1}{n} \sum_{d=1}^n \( \hat{w}_{i,q+d} - \hat{g}_{i,q+d} (\btheta_i) \) ^2$ with $n= N-q$ samples and $\lambda_n$ is a tuning parameter defined in Theorem \ref{thm-2}.
We note that the lag order $q$ is allowed to increase with $N$.
To analyze its asymptotic behaviors, we need the following technical conditions.

\begin{assumption}\label{assumption3}~
\begin{enumerate}
\item [(a)] $\bW_d= (\bw_d ^{\top}, \ldots, \bw_{d-q+1} ^{\top}) ^{\top} $ is strictly stationary, and the spectral radius of $\bA$ is less than one, where 
$$
\bA= \begin{pmatrix}
\bbeta_1 & \bbeta_2 & \ldots &\bbeta_{q-1} & \bbeta_q \\ 
 \bI_p &0  & \cdots&  0 &0\\ 
0 & \bI_p  &0 & \cdots  & 0 \\
\vdots & & \ddots&0 &0 \\
0&0&0& \bI_p&0 
\end{pmatrix}.
$$
 
\item [(b)] The number of nonzero elements of $\btheta_{0,i}$ is bounded by $s_{\beta} \geq1$, where  $\btheta_{0,i}$ is the true parameter for the $i$th weight. 

\item [(c)] The process $\bW_d$ is $\alpha$-mixing, and the $\alpha$ mixing coefficients satisfy $\alpha (k) = O( c^k ) $  for some $c \in (0,1)$.

\item [(d)] $\max_d \|\hat{\bw}_d - \bw_d\|_{\max} \leq C  s_p \tau_m ^{1-\delta}$. 

\item [(e)] Define  $\mathcal U_{i} = \left\{ \bu \in \mathbb{R}^{qp+1}:\text{ }\left\| \bu_{S_{i}^{c}} \right\|_{1} \leq 3\left\| \bu_{S_{i}}\right\|_{1}\right\}$, where
$S_{i} = \{j: \text{ jth element of } \btheta_{0,i} \neq 0 \}$, 
$\bu_{S_{i}}$ is the subvector obtained by stacking $\left\{\bu_{j}: \text{ } j \in S_{i} \right\}$, and
$\bu_{S_{i}^{c}}$ is the subvector obtained by stacking $\left\{\bu_{j}: \text{ } j \in S_{i}^{c} \right\}$.
Then, there exists a constant $\kappa>0$ such that the following inequality holds for some $D \geq 48s_{\beta}\lambda_n / \kappa$ and $1 \leq i \leq p$, where the specific value of $\lambda_n$ is given in Theorem \ref{thm-2}:
\begin{equation*}
\inf\{ \bu^{\top}\nabla ^{2}{\mathcal{L}}_{n,i}(\btheta_i) \bu: \text{ } \bu \in \mathcal U_{i}, \text{ } \left\| \bu\right\| _{2}= 1,\text{ } \left\| \btheta_i -\btheta_{0,i}\right\| _{1}\leq D
\} \geq \kappa.
\end{equation*}
\end{enumerate}
\end{assumption}

\begin{remark}
Assumption \ref{assumption3}(a) is the strictly stationary and stable condition for the VAR(1) model for $\bW_d$, which is a first-order representation of \eqref{eq:g-AR}.
For the model \eqref{eq:DRMVPvol}, the strictly stationarity holds if $\bZ_{t}$ is a strictly stationary process and the spectral radius of $\bA$ is less than one.
Assumption \ref{assumption3}(c) is satisfied with an exponentially decaying absolute summation condition on coefficients of MA$(\infty)$ representations of $\bW_{d}$ (see Assumption (A2) in \citet{masini2022regularized} and Theorem 2.1 in \citet{chanda1974strong}), which restricts the coefficients $\bbeta_j$'s.
Assumption \ref{assumption3}(d) is satisfied with high probability by Proposition \ref{thm1}. 
In contrast, when we employ other estimation procedures for the MVP weights, as long as Assumption \ref{assumption3}(d)-type condition is satisfied, we can obtain a similar result. 
That is, as long as the realized non-normalized MVP estimator performs well, the proposed estimation method works.
Finally, Assumption \ref{assumption3}(e) is the eigenvalue conditions for the Hessian matrix $\nabla ^{2}{\mathcal{L}}_{n,i}(\btheta_i) $.
This is called the localized restricted eigenvalue (LRE) condition \citep{fan2018lamm, shin2021factor, sun2020adaptive}, which implies strictly positive restricted eigenvalues over a local neighborhood.
\end{remark}

  The following theorem establishes the asymptotic convergence rate for the LASSO estimation procedure.

\begin{theorem} \label{thm-2}
  Suppose that $N$ is greater than the lag order $q$ such that the number of usable observations $n=N-q$ diverges as $N$ increases.
  Under Assumptions \ref{assumption1}--\ref{assumption3}, take $\lambda_n \geq C (s_{\beta} s_p \tau_m ^{1-\delta} +    n ^{-1/2} \sqrt{ \log (p \vee q)  } ) $ for some large constant $C$, which goes to zero as $p$ and $n$ go to infinity.
  Then, we have 
  \begin{equation} \label{thm2-result}
	  \max_i \|\hat{\btheta}_i - \btheta_{0,i} \|_1 \leq C   \frac{s_{\beta}  \lambda_n }{\kappa} \quad \text{and} \quad  \max_i \|\hat{\btheta}_i - \btheta_{0,i} \|_2 \leq C   \frac{s_{\beta}^{1/2} \lambda_n }{\kappa}
  \end{equation}
  with probability at least $1- (p \vee q)^{-c}$ for any given positive constant $c$.
\end{theorem}
  
\begin{remark}
  Theorem \ref{thm-2} shows that the LASSO estimator has the convergence rate $\lambda_n$ with $s_{\beta}$ term. 
  When we consider exact sparsity for the inverse matrix ($\delta=0$), we have the convergence rate $m^{-1/4} +n^{-1/2}$ with additional $s_p$ and $s_{\beta}$ up to log order. 
  The term $m^{-1/4}$ is the cost to estimate the unobserved non-normalized MVP using the high-frequency observations. 
  The term $n^{-1/2}$ is the usual convergence rate for estimating low-frequency models with the $n$ period.
  Given the lag order $q$, the total number of variables for the LASSO estimator is $pq$.
  Therefore, the lag order $q$ and the number of assets $p$ affect the convergence rate $\lambda_n$ of the LASSO estimator.
\end{remark}

The main purpose of this paper is to predict the future MVP.
To do this, we estimate the conditional MVP weight as $\hat{ \bar{\bw}}_{n+1} = \frac{\hat{\bg}_{n+1} ( \hat{\btheta}_i) }{\b1 ^{\top} \hat{\bg}_{n+1} ( \hat{\btheta}_i)}$.
In the following theorem, we investigate its asymptotic properties.

\begin{theorem}\label{thm-3}
  Under the assumptions in Theorem \ref{thm-2}, we have
  \begin{equation}\label{thm3-result1}
	  \|\hat{\bg}_{n+1} ( \hat{\btheta})  - \bg_{n+1}( \btheta_{0}) \|_{\max} \leq  C   \frac{s_{\beta}  \lambda_n }{\kappa}
  \end{equation}
  and 
  \begin{equation}\label{thm3-result2}
    \| \hat{ \bar{\bw}}_{n+1}  - E \(\bar{\bw}_{n+1} \middle | \FF_n \) \|_1 \leq C  \( \frac{s_{\beta}  \lambda_n }{\kappa} + \left \|\frac{\bg_{n+1} (\btheta)}{\b1^{\top} \bg_{n+1} (\btheta)}  \right  \|_{1}    p^{-1/2} \)
  \end{equation}
  with probability at least $1- (p \lor q)^{-c}$ for any given positive constant $c$.
\end{theorem}

Theorem \ref{thm-3} indicates that the proposed conditional expected non-normalized MVP estimator $\hat{\bg}_{n+1} ( \hat{\btheta}) $ can consistently estimate the future non-normalized MVP with the $s_{\beta}  \lambda_n$ order.
Furthermore, the conditional MVP weight estimator has the convergence rate $ s_{\beta}  \lambda_n   + \left \|\frac{\bg_{n+1} (\btheta)}{\b1^{\top} \bg_{n+1} (\btheta)}  \right  \|_{1}    p^{-1/2}$.
The first term  $ s_{\beta}  \lambda_n$ is the cost to estimate unobserved future non-normalized MVP using the high-frequency and low-frequency observations.
The second term $\left \|\frac{\bg_{n+1} (\btheta)}{\b1^{\top} \bg_{n+1} (\btheta)}  \right  \|_{1}    p^{-1/2}$ is the approximation error. 
From Theorem \ref{thm-3}, we can conclude that the proposed estimation method can consistently estimate the future MVP.

\subsection{Discussion on Price Jumps}\label{JumpDiffusionProcess}

In financial practice, we often observe jumps.
In the presence of price jumps, quadratic covariation can be decomposed into continuous and jump components, and the product of its inverse and one vector is the non-normalized MVP in the ex-post sense.
On the other hand, several empirical studies reported that the volatility dynamics can be better explained by decomposing quadratic covariation into its continuous and jump components, since jumps are associated with news announcements and have less persistent features \citep{andersen2007roughing,barndorff2006econometrics,corsi2010threshold,lee2008jumps}.
In a similar spirit, employing MVP using integrated volatility rather than quadratic covariation may help capture the conditional MVP dynamics well.
To check this, we conduct an empirical study to compare integrated volatility and quadratic covariation and report the results in the online Appendix \ref{compare-IQ}. 
We find that employing integrated volatility shows better performance.
Details can be found in the online Appendix \ref{compare-IQ}.

To estimate integrated volatility in the presence of price jumps, we first detect jumps from observed stock log-return data and estimate the integrated volatility.
For example,  \citet{fan2007multi} and \citet{zhang2016jump} employed the wavelet method to identify the jumps based on noisy high-frequency data.
\citet{mancini2004estimation} studied a threshold method for jump detection and presented the order of an optimal threshold, and \citet{davies2018data} further examined a data-driven threshold method.
They demonstrated that the estimator of jump variation has the convergence rate of $m^{-1/4}$, which further helps the estimator of integrated volatility to achieve the optimal convergence rate of $m^{-1/4}$. 
In this paper, we employ the jump-robust pre-averaging realized volatility matrix estimator (JPRVM) \citep{ait2016increased} and describe the procedure in the online Appendix \ref{SEC:jprvm}. 
Then, we can show that this estimator satisfies Assumption \ref{assumption2}(a) under the sub-Gaussian condition, and with it, we can apply the proposed estimation procedure and obtain the same result.

\subsection{Choice of the Tuning Parameters}\label{Section-TuningParameter}
To apply the proposed estimators, we need to choose the tuning parameters $\tau_{m}$ and $\lambda_n$ for the CLIME estimator and the LASSO estimator, respectively.
For the CLIME estimator, we estimate it by varying $C_{\tau}$ of $\tau_m = C_{\tau} m^{-1/4} \sqrt{\log \left( p \vee N \right) }$, and we select $C_{\tau}$ among 100 logarithmically spaced points ranging from $10^{-6}$ to $10$, which minimizes the likelihood loss, as follows:
\begin{eqnarray*}
	&& \hat{\bOmega}_{d,\tau} = \argmin_{\bOmega} \norm{\bOmega}_{1} \text{ s.t. }  \norm{\hat{\bGamma}_{d} \bOmega  - \bI }_{\max} \leq \tau, \\
	&& \hat{\bOmega}_{d} = \hat{\bOmega}_{d,\tau^{*}}, \text{ where } \tau^{*} = \argminU_{\tau \in \left\lbrace \tau_m | C_{\tau} \in [10^{-6},10] \right\rbrace} \langle \hat{\bOmega}_{d,\tau}, \hat{\bGamma}_{d} \rangle - \log\det ( \hat{\bOmega}_{d,\tau} ) .
\end{eqnarray*}
For the LASSO estimator, we need to choose the tuning parameters $\lambda_n$.
The estimation of the DR-MVP model typically faces the small-n-large-P situation \citep{chen2008extended}.
For example, in our empirical study (Section 5), we have $n=252-50=202$ and $P=200\times50=10000$ for the estimation of the DR-MVP model with an AR$(50)$ specification.
Therefore, we select the tuning parameter $\lambda_n$ from 100 logarithmically spaced points ranging from $10^{-6}$ to $10$, which minimizes the extended Bayesian information criterion (EBIC) \citep{chen2008extended,chen2012extended} of the LASSO estimation, where
\begin{equation*}
	\text{EBIC} = n\log(\mathcal{L}_{n,i}(\btheta_{i})) + \norm{\btheta_{i}}_{0} \log(n)   + 2 \norm{\btheta_{i}}_{0} \gamma \log P ,
\end{equation*}
$\mathcal{L}_{n,i} = \frac{1}{n} \sum_{d=1}^{n} (\hat{w}_{i,d} - \hat{g}_{i,d}(\btheta_{i}))^{2}$, $\gamma=0.5$, as \citet{chen2012extended} suggested, and $P$ represents the number of covariates under consideration.

\section{Simulation Study}\label{SEC-4}
We conducted simulations to show that the proposed methodology has good finite sample performance and compared the proposed method with other existing methods.
In Section \ref{SEC-5}, we find that the heterogeneous autoregressive (HAR) specification for the DR-MVP model can help explain the MVP dynamics.
To reflect this feature, we considered the HAR-type VAR$(22)$ structure for the DR-MVP model, which is
\begin{equation}\label{eq:simulation-HARspec}
  \bw_{d} = \bbeta_{0} + \bbeta_{1}  \bar{\bw}_{d}^{(1)} + \bbeta_{5}  \bar{\bw}_{d}^{(5)} + \bbeta_{22}  \bar{\bw}_{d}^{(22)} + \bepsilon_{d}
  ,
\end{equation}
where $\bar{\bw}_{d}^{(k)}=k^{-1} \sum_{j=1}^{k} \bw_{d-j}$ is the average level of MVP weights over the previous $k$ days.
The true log-prices for the $p$ assets follow the jump-diffusion process
\begin{equation*}
  d\bX_{t} = \bsigma_{t} d\bB_{t} + \bJ_{t}  d\bLambda_t
  ,
\end{equation*}
where $\bB_{t}$ is a $p$-dimensional standard Brownian motion, $\bSigma_{t} = \bsigma_{t} \bsigma_{t}^{\top}$ is the instantaneous volatility process, the jump sizes $\bJ_t$ obey the independent normal distribution $N(0.05,0.005^2)$ with randomly determined signs, and $\bLambda_t$ is a $p$-dimensional Poisson process with intensity 5.
The instantaneous volatility matrix $\bSigma_{t}$ follows the data-generating process introduced in the online Appendix \ref{SEC:sample-model}, so the DR-MVP model in \eqref{eq:simulation-HARspec} is satisfied.
For each simulation process, we generated high-frequency data with $m=23400$  for 500 consecutive days.
We used the subsampled log-prices of the last $N=125, 250, 500$ days with high-frequency observations $m=2340,7800,23400$ per day and repeated the simulation procedure 100 times.

To estimate the integrated volatility matrix, we utilized the POET procedure \citep{fan2013large}.
Specifically, the input integrated volatility matrix for the POET procedure is estimated by the JPRVM estimator in the online Appendix \ref{SEC:jprvm}.
We then took a hard thresholding function with threshold level $\sqrt{\log p/m^{1/2}} + 1/\sqrt{p}$ for the idiosyncratic volatility matrix.

\begin{figure}[!ht] 
	\centering
  \includegraphics[width = 0.8\textwidth]{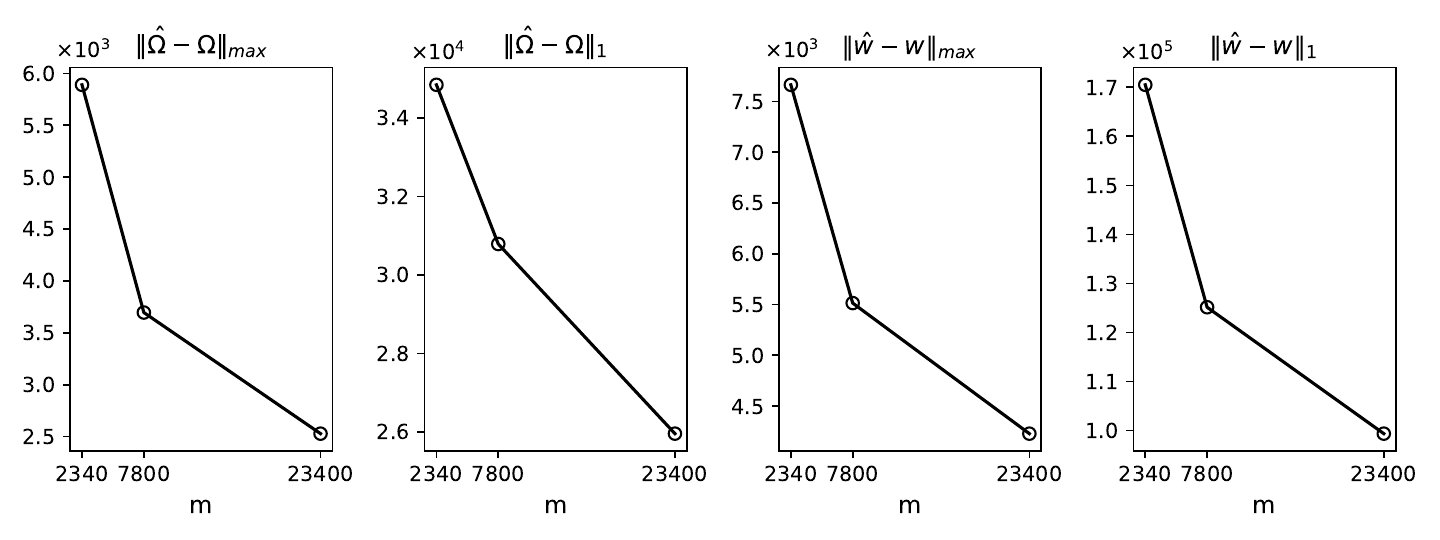} 
	\caption{Mean of matrix max and $\ell_1$-norms of ${\hat{\bOmega}_{d} - \bOmega_{d}}$ and mean max and $\ell_1$-norms of $\hat{\bw}_{d} - \bw_{d}$ for the CLIME and realized non-normalized MVP estimators with $m=2340,7800,23400$.}
	\label{fig:NonParametric}
\end{figure}
 
Figure \ref{fig:NonParametric} draws the mean of matrix  $\ell_1$-norms and max norms of ${\hat{\bOmega}_{d} - \bOmega_{d}}$ and mean max and $\ell_1$-norms of $\hat{\bw}_{d} - \bw_{d}$ for the CLIME and realized non-normalized MVP estimators with $m=2340,7800,23400$.
From Figure \ref{fig:NonParametric}, we find that the mean estimation errors of non-parametric estimators decrease as the number of high-frequency observations increases.
These results support the theoretical results derived in Proposition \ref{thm1}.

\begin{figure}[!ht] 
	\centering
	\includegraphics[width = 0.75\textwidth]{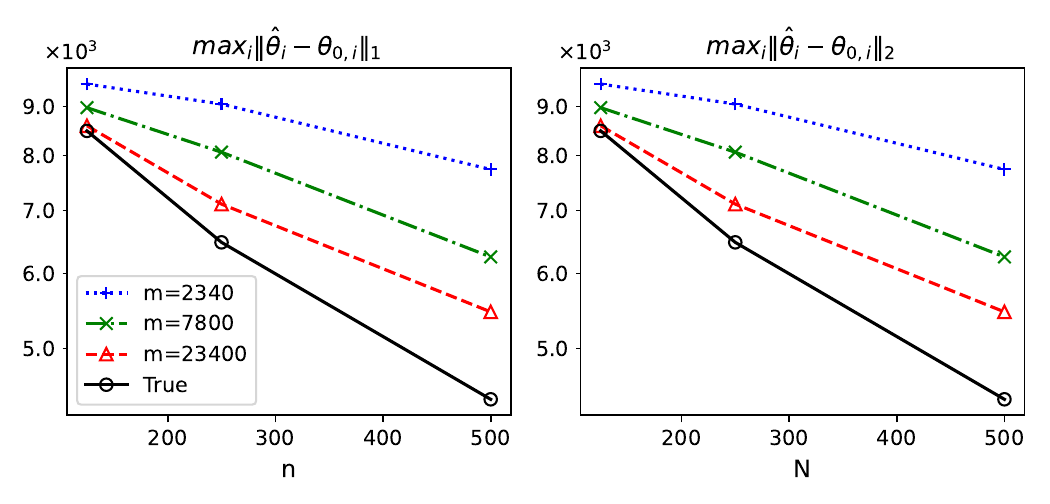} 
	\caption{Mean errors of $\max_{i} \lVert \hat{\btheta}_{i} - \btheta_{0,i} \rVert_{1}$ and $\max_{i} \lVert \hat{\btheta}_{i} - \btheta_{0,i} \rVert_{2}$ for the proposed DR-MVP estimator with $N=125,250,500$ and $m=2340,7800,23400$, and the true MVP.}
	\label{fig:ParameterEstimation}
\end{figure}

We then checked the finite sample performances of the DR-MVP model.
We first estimated the model parameters by the LASSO  for $N=125,250,500$ and $m=2340,7800,23400$, and the cases employing the true MVPs.
Figure \ref{fig:ParameterEstimation} depicts the mean errors of $\max_{i} \lVert \hat{\btheta}_{i} - \btheta_{0,i} \rVert_{1}$ and $\max_{i} \lVert \hat{\btheta}_{i} - \btheta_{0,i} \rVert_{2}$ for the DR-MVP estimator.
From Figure \ref{fig:ParameterEstimation}, we find that the mean errors decrease as the number of high- or low-frequency observations increases.
These results support the theoretical findings in Theorem \ref{thm-2}.

The main purpose of this paper is to predict the future MVP.
Thus, we investigated the out-of-sample performance of estimating the one-day-ahead conditional non-normalized and normalized MVP $\bg_{d+1}(\btheta_{0})$ and $\mathbb{E}\left[ \bar{\bw}_{d+1} | \mathcal{F}_{d} \right]$, where we calculated $\mathbb{E}\left[ \bar{\bw}_{d+1} | \mathcal{F}_{d} \right]$ by the Monte Carlo method.
For comparisons, we considered one additional DR-MVP model that utilizes the simple HAR model (HAR) \citep{corsi2009simple} for estimating the future non-normalized MVP weights, and HAR employs the OLS procedure to estimate parameters.
We also consider a non-parametric estimator that only utilizes the latest CLIME estimator, such as  $\hat{\bOmega}_{d} \b1$ and $\frac{\hat{\bOmega}_{d} \b1}{\b1^{\top} \hat{\bOmega}_{d} \b1} $.
The CLIME estimator is based on the martingale assumption for the non-normalized MVP weights.
Finally, we employed the parametric estimator FIVAR \citep{shin2021factor}, which can account for the factor and idiosyncratic volatility dynamics.
With regard to the FIVAR estimator, we first obtained the estimated conditional volatility matrix $\hat{\bGamma}_{d+1}^{F}$ by the FIVAR estimator, and we calculated the non-normalized and normalized MVPs as
$(\hat{\bGamma}_{d+1}^{F})^{-1} \b1$ and $\frac{(\hat{\bGamma}_{d+1}^{F})^{-1} \b1}{\b1^{\top} (\hat{\bGamma}_{d+1}^{F})^{-1} \b1}$, respectively.
To estimate the one-day-ahead FIVAR volatility matrix, we followed the same procedure suggested in \citet{shin2021factor}.

\begin{figure}[!ht] 
	\centering
	\includegraphics[width = 0.85\textwidth]{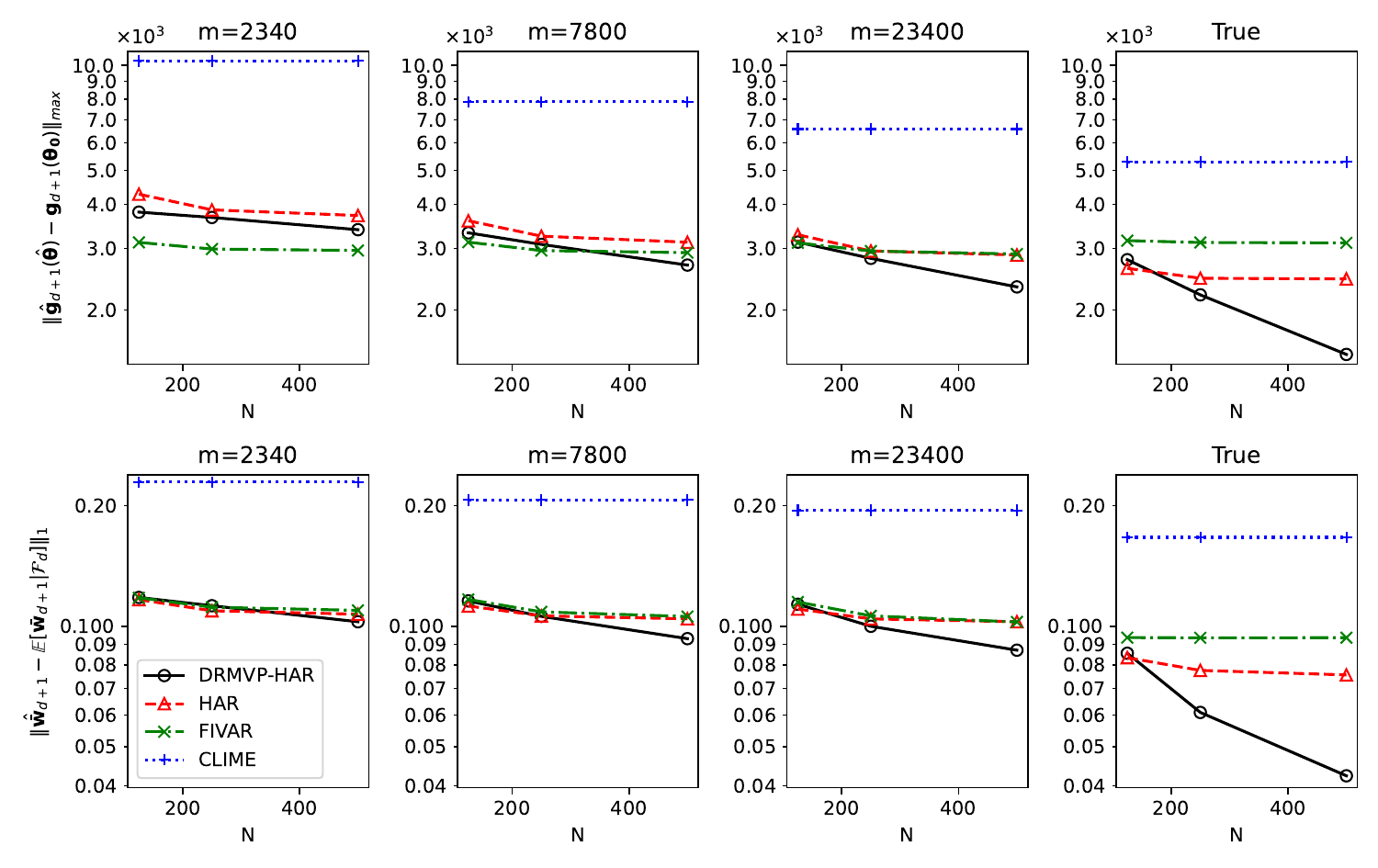} 
	\caption{Mean of $\lVert \hat{\bg}_{d+1}(\hat{\btheta}) - \bg_{d+1} (\btheta_{0}) \rVert_{\max}$ and $\lVert \hat{\bar{\bw}}_{d+1} - \mathbb{E}\left[ \bar{\bw}_{d+1} | \mathcal{F}_{d} \right] \rVert_{1}$ for $N=125,250,500$ and $m=2340,7800,23400$, along with cases employing the true values in place of nonparametric estimators.}
	\label{fig:PredictSimulation}
\end{figure}

To provide numerical support for Theorem \ref{thm-3}, Figure \ref{fig:PredictSimulation} draws the mean of $\lVert \hat{\bg}_{d+1}(\hat{\btheta}) - \bg_{d+1} (\btheta_{0}) \rVert_{\max}$ and $\lVert \hat{\bar{\bw}}_{d+1} - \mathbb{E}\left[ \bar{\bw}_{d+1} | \mathcal{F}_{d} \right] \rVert_{1}$ for $N=125,250,500$ and $m=2340,7800,23400$, along with cases employing the true values in place of nonparametric estimators.
From Figure \ref{fig:PredictSimulation}, we find that the parametric models perform better than the nonparametric model CLIME.
When comparing the parametric models, the DRMVP-HAR model performs better than others for higher $m$ and $n$, while the FIVAR model performs better in terms of $\lVert \hat{\bg}_{d+1}(\hat{\btheta}) - \bg_{d+1} (\btheta_{0}) \rVert_{\max}$ for $m=2340$.
This implies that we need sufficient high-frequency observations to utilize the estimated MVP as inputs of the DR-MVP models.
Moreover, we find that the mean prediction errors of the conditional non-normalized and normalized MVPs decrease as the number of high- or low-frequency observations increases.
These results support the theoretical results in Theorem \ref{thm-3}.
We note that additional prediction error analyses for different norms can be found in the online  Appendix \ref{SEC:appendix-simulation}.

\begin{figure}[!ht] 
	\centering
	\includegraphics[width = 0.85\textwidth]{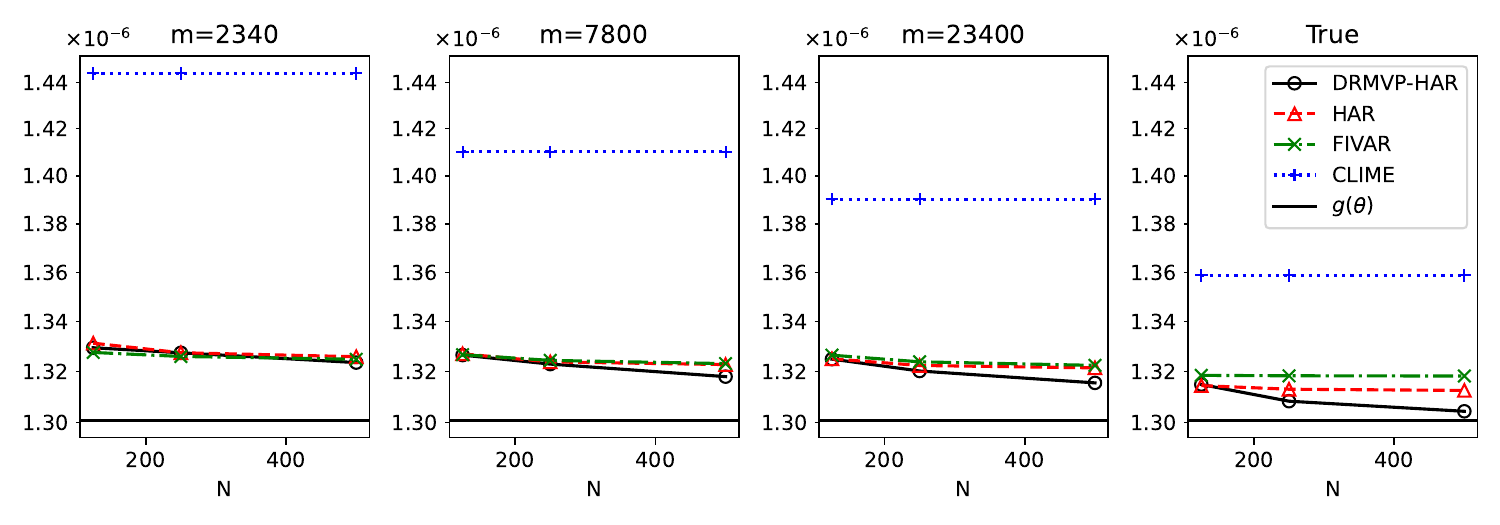} 
	\caption{Mean out-of-sample normalized portfolio risk for the DR-MVP, FIVAR, and CLIME models with $N=125,250,500$ and $m=2340,7800,23400$, along with cases employing the true values in place of nonparametric estimators.}
	\label{fig:SimulationRisk}
\end{figure}

We further investigated the out-of-sample risk of the minimum variance portfolio allocation problem.
The out-of-sample risk of a normalized portfolio $\hat{\bar{\bw}}_{d+1}$ is calculated as $ R = \hat{\bar{\bw}}_{d+1}^{\top} \bGamma_{d+1} \hat{\bar{\bw}}_{d+1}$,
where $\bGamma_{d+1}$ is the one-day-ahead integrated volatility matrix.
Figure \ref{fig:SimulationRisk} depicts the mean out-of-sample normalized portfolio risk for the DR-MVP, FIVAR, and CLIME models with $N=125,250,500$ and $m=2340,7800,23400$, and the cases employing the true values in place of nonparametric estimators.
The horizontal black solid line in Figure \ref{fig:SimulationRisk} indicates the mean out-of-sample normalized portfolio risk that can be achieved with the true $\bg_{d+1}(\btheta)$.
Figure \ref{fig:SimulationRisk} shows that the DR-MVP model performs best among the benchmarks, except for the small $m$ case ($m=2340$).
This is because the DR-MVP model can directly capture the MVP dynamics.
On the other hand, the mean out-of-sample risk for the DR-MVP, FIVAR, and CLIME models decreases as the number of high- or low-frequency observations increases.
This may be because, as the number of high- or low-frequency observations increases, we can obtain more accurate input volatility matrix estimator and parametric estimations, respectively.
We end this section by remarking that the proposed DR-MVP model shows comparable performance when the true model is misspecified.
To assess the robustness of the proposed methodology when the true model is misspecified, we conducted an additional analysis assuming that the data-generating process follows a different specification, namely the FIVAR data generating process \citep{shin2021factor}.
The full methodology and results of this robustness analysis are presented in the online Appendix B.

\section{Empirical study}\label{SEC-5}
In this section, we apply the proposed DR-MVP model to real high-frequency stock trading data.
We obtained high-frequency data from January 2016 to December 2017, 503 trading days in total, for the top 200 large trading volume stocks among the components of the S\&P 500 from the TAQ database in the Wharton Research Data Service (WRDS) system.
In order to calculate the covariances between different assets, we employed the pairwise-refresh time scheme proposed by \citet{fan2012vast}  to obtain synchronized samples.
The average values of daily means, medians, standard deviations, minimums, and maximums of the number of daily synchronized samples are 6283, 5474, 4030, 1968, and 225180, respectively.

To employ the proposed DR-MVP model, we first needed to obtain the inverse matrix estimators of the realized volatility matrix estimators by the CLIME estimation procedure.
For the input volatility matrix, we used the JPRVM estimator defined in the online Appendix \ref{SEC:jprvm} to the synchronized intraday log-prices for each pair of stocks.
We then employed the POET procedure introduced by \citet{fan2013large, fan2016incorporating} and used the global industry classification standard (GICS) for the threshold level for the idiosyncratic part \citep{fan2016incorporating}.
Specifically, we kept the idiosyncratic volatilities within the same sector, but set zero for others.
Then, with the POET estimator, we applied the CLIME estimator to obtain the inverse matrix estimations $\hat{\Omega}_{d}$ and the MVPs $\hat{\bw}_{d} = \hat{\Omega}_{d} \b1$ with the tuning parameters $\tau_{n}$, chosen by the procedure  in Section \ref{Section-TuningParameter}.

We then estimated $\hat{\bg}_{d+1} (\hat{\btheta})$ by LASSO with the tuning parameters $\lambda_{n}$, chosen by the procedure in Section \ref{Section-TuningParameter}.
For the explanatory variables of $\hat{\bg}_{d+1} (\btheta)$, we considered two cases, such as 50 lagged realized MVPs (DRMVP-50) and HAR-type realized MVPs (DRMVP-HAR).
The latter can reproduce relatively long memory persistence, such as AR(22),  while remaining parsimonious.
Specifically, we assumed the VAR structure for $\hat{\bg}_{d+1} (\btheta)$ with HAR-type realized MVPs as follows:
\begin{equation}\label{eq:LASSO-HAR}
	\hat{\bg}_{d+1} (\btheta) = \bbeta_{0} + \sum_{h\in\{1,5,22\}} \bbeta_{h} \hat{\bw}_{d}^{(h)}
  ,
\end{equation}
where $\hat{\bw}_{d}^{(h)} = \frac{1}{h} \sum_{j=d-h+1}^{d} \hat{\bw}_{j}$. 
We used the rolling window scheme to predict the one-day-ahead non-normalized MVPs, where the in-sample period was 252 days.

To check if the proposed model had captured the MVP dynamics well, we investigated the out-of-sample performance of the minimum variance portfolio allocation problem.
We computed the out-of-sample annualized portfolio risk of the normalized portfolios for a given period with $d$ days.
Furthermore, to assess how close the out-of-sample portfolio risk is to the ex-post minimum variance portfolio risk, we calculated the mean relative portfolio risks.
Specifically, the annualized risk and the mean relative risk are calculated as follows:
\begin{equation*}
  \text{Annualized risk} = \sqrt{\frac{252}{d} \sum_{k=1}^{d} \sum_{i=1}^{39}  r_{k,i}(\hat{\bar{\bw}}_{k})^{2} } \quad \text{and }  \text{Mean relative risk} = \frac{1}{d} \sum_{k=1}^{d} \frac{\sum_{i=1}^{39} r_{k,i}(\hat{\bar{\bw}}_{k})^{2}}{\sum_{i=1}^{39} r_{k,i}({\bar{\bw}}_{k})^{2}} 
  ,
\end{equation*}
where $\hat{\bar{\bw}}_{k}$ is a normalized portfolio and $r_{k,i}(\hat{\bar{\bw}}_{k}) = \hat{\bar{\bw}}_{k}^{\top} (\bY(k-1+\frac{i}{39} ) - \bY(k - 1+\frac{i-1}{39} ))$ is the 10-minute portfolio log-return and $\bar{\bw}_{k}$ is the ex-post normalized minimum variance portfolio obtained from the CLIME estimator.
Finally, we computed the mean $\ell_2$-norm of the difference between the estimated out-of-sample normalized MVP and the ex-post normalized MVP, $\sum_{k=1}^{d} \Vert \hat{\bar{\bw}}_{k} - {\bar{\bw}}_{k}\Vert_{2} /d$.
We used two different out-of-sample periods, namely from day 253 to day 378 and from day 379 to day 503, denoted by Period 1 and Period 2, respectively. 
We considered two benchmarks based on the CLIME and FIVAR estimators and one additional DR-MVP model that utilizes the simple HAR model (HAR), defined in Section \ref{SEC-4}.
For example, HAR follows the proposed DR-MVP model, but the off-diagonal elements of the $\(\bbeta_{i,h}\)_{h=1,5,22}$ are restricted to zero and $\hat{\btheta_i}= \arg \min_{\btheta_i}   \mathcal{L}_{n,i} (\btheta_i)$.
All the benchmark portfolios were normalized to measure the out-of-sample portfolio risks.

\begin{table}[!ht]
\caption{Out-of-sample annualized and relative risks of the normalized MVP with the DR-MVP, HAR, FIVAR, and CLIME models for Period 1, Period 2, and the whole period.}
\label{Table-risks}
\centering
\begin{tabular}{lrrrrrlrrrrr}
\hline
       & \multicolumn{5}{c}{Annualized risks}                                                                                                            \\ \cline{2-6}
Period & \multicolumn{1}{c}{DRMVP-50} & \multicolumn{1}{c}{DRMVP-HAR} & \multicolumn{1}{c}{HAR} & \multicolumn{1}{c}{FIVAR} & \multicolumn{1}{c}{CLIME}  \\ \hline
1      & 4.667\%                      & 4.660\%                       & 4.512\%                 & 4.923\%                   & 4.801\%                    \\
2      & 4.199\%                      & 4.182\%                       & 4.124\%                 & 4.447\%                   & 4.333\%                    \\
Whole  & 4.438\%                      & 4.427\%                       & 4.322\%                 & 4.690\%                   & 4.572\%                    \\ \hline
       & \multicolumn{5}{c}{Mean relative risks}                                                                                                         \\ \cline{2-6}
Period & \multicolumn{1}{c}{DRMVP-50} & \multicolumn{1}{c}{DRMVP-HAR} & \multicolumn{1}{c}{HAR} & \multicolumn{1}{c}{FIVAR} & \multicolumn{1}{c}{CLIME}  \\ \hline
1      & 1.339                        & 1.337                         & 1.301                   & 1.434                     & 1.379                      \\
2      & 1.358                        & 1.351                         & 1.323                   & 1.479                     & 1.412                      \\
Whole  & 1.348                        & 1.344                         & 1.312                   & 1.457                     & 1.396                      \\ \hline
\end{tabular}
\end{table}

Table \ref{Table-risks} reports the out-of-sample annualized risks and relative risks of the normalized MVP with the DRMVP-50, DRMVP-HAR, HAR, FIVAR, and CLIME models.
From Table \ref{Table-risks}, we find that the DR-MVP-based models (DRMVP-50, DRMVP-HAR, and HAR) show the best performance. 
The HAR model is slightly better than the DRMVP-50 and DRMVP-HAR models.
This may be because the DR-MVP models help capture the MVP dynamics, while the estimation error of the HAR model parameter is smaller than that of the LASSO-based parameter.
That is, the sparsity structure on the HAR model parameters may explain the MVP dynamics well.
However, the difference is relatively small; thus, the LASSO procedure can also capture the MVP dynamics.
On the other hand, the model using the CLIME estimator to obtain the inverse matrix performs better than one using the direct inverse operation, such as the FIVAR model.
This may be because accommodating the sparsity of the inverse matrix helps account for the MVP weights.

\begin{table}[!ht]
\caption{Average rank of daily 10-minute portfolio risks for the DRMVP-50, DRMVP-HAR, HAR, FIVAR, and CLIME models for Period 1, Period 2, and the whole period. In the parentheses, we report the number of those ranked first among the models.}
\label{Table-rank}
\centering
\begin{tabular}{lrrrrr}
\hline
Period & \multicolumn{1}{c}{DRMVP-50} & \multicolumn{1}{c}{DRMVP-HAR} & \multicolumn{1}{c}{HAR} & \multicolumn{1}{c}{FIVAR} & \multicolumn{1}{c}{CLIME}  \\ \hline
1      & 3.192 (20)                   & 2.800 (11)                    & 2.520 (45)              & 2.856 (23)                & 3.632 (26)                 \\
2      & 3.238 (15)                   & 2.524 (17)                    & 2.167 (45)              & 3.151 (21)                & 3.921 (28)                 \\
Whole  & 3.215 (35)                   & 2.661 (28)                    & 2.343 (90)              & 3.004 (44)                & 3.777 (54)                 \\ \hline
\end{tabular}
\end{table}

In addition to evaluating annualized and relative risks, we computed daily risk ranks for the DRMVP-50, DRMVP-HAR, HAR, FIVAR, and CLIME models.
Since the rank measure effectively mitigates the undue effect of outlying risks, which can distort annualized and relative risk measures, the rank measure serves as a metric to evaluate whether the proposed model consistently performs better than the other models.
Table \ref{Table-rank} reports the average rank and the number of those ranked first of daily 10-minute portfolio risks for the DRMVP-50, DRMVP-HAR, HAR, FIVAR, and CLIME models.
From Table \ref{Table-rank}, we find that the HAR model consistently shows the best performance.
Meanwhile, the FIVAR model has better average ranks than the DRMVP-50 and CLIME models, whereas the DRMVP-50 and CLIME have smaller annualized risks and mean relative risks.
One possible explanation is that the CLIME-based weights are distinct from the FIVAR weights, and the CLIME-based estimators are likely to have similar behavior in terms of the 10-minute portfolio risks.
In this situation, the average rank can be more advantageous for FIVAR than DRMVP-50 and CLIME that show consistently worse performance among the CLIME-based estimators.

\begin{table}[!ht]
\caption{Mean $\ell_2$-norm of the difference between the ex-post normalized MVP obtained from the CLIME estimator and the estimated normalized MVP using the DRMVP-50, DRMVP-HAR, HAR, FIVAR, and CLIME models for Period 1, Period 2, and the whole period.}
\label{Table-l2}
\centering
\begin{tabular}{lrrrrr}
\hline
Period & \multicolumn{1}{c}{DRMVP-50} & \multicolumn{1}{c}{DRMVP-HAR} & \multicolumn{1}{c}{HAR} & \multicolumn{1}{c}{FIVAR} & \multicolumn{1}{c}{CLIME}  \\ \hline
1      & 0.154                        & 0.154                         & 0.148                   & 0.169                     & 0.191                      \\
2      & 0.134                        & 0.133                         & 0.127                   & 0.140                     & 0.166                      \\
Whole  & 0.144                        & 0.143                         & 0.138                   & 0.154                     & 0.179                      \\ \hline
\end{tabular}
\end{table}

Table \ref{Table-l2} reports the mean $\ell_2$-norm of the difference between the ex-post normalized MVP obtained from the CLIME estimator and the estimated normalized MVP using the DRMVP-50, DRMVP-HAR, HAR, FIVAR, and CLIME models for Period 1, Period 2, and the whole period.
From Table \ref{Table-l2}, we find that the DR-MVP-based models perform better than the other benchmarks.
The HAR model shows the best performance among the DR-MVP-based models.
The mean of $\ell_2$-norm of the CLIME model is greater than that of the FIVAR model, although the annualized risks and the mean relative risks of the CLIME are less than that of the FIVAR.
This may be due to the fact that the FIVAR model captures the linear dynamics of both factor and idiosyncratic volatilities using past volatility matrices.
Therefore, by Taylor's expansion,  the FIVAR model is able to partially approximate the dynamics of the MVP.
From these results, we can conjecture that the proposed DR-MVP model can capture the MVP dynamics.

To compare the predictive accuracy among the DRMVP-50, DRMVP-HAR, HAR, FIVAR, and CLIME models, we conducted Diebold-Mariano (DM) tests \citep{diebold2002comparing} for the 10-minute portfolio risks, relative risks, and $\ell_2$-norms.
We compared the DRMVP-HAR model with other models as follows:
\begin{equation*}
  H_0 : \mathbb{E}\left[ R_{i}^{*} \right] = \mathbb{E}\left[ R_{i} \right]  \quad \text{vs.} \quad H_1 : \mathbb{E}\left[ R_{i}^{*} \right]  < \mathbb{E}\left[ R_{i} \right] 
  ,
\end{equation*}
where $R_{i}^{*}$ and $R_{i}$ are one of the 10-minute portfolio risks, relative risks, or $\ell_2$-norms for the DRMVP-HAR model and the other models, respectively.
 Table \ref{Table-dm-relrisk} reports the $p$-values of the DM tests based on 10-minute portfolio risks, relative risks, and $\ell_2$-norm for the DRMVP-HAR model versus one of the DRMVP-50, HAR, FIVAR, and CLIME models for Period 1, Period 2, and the whole period.
 From Table \ref{Table-dm-relrisk}, we can find that the $p$-values of DM tests based on 10-minute portfolio risks, relative risks, and $\ell_2$-norm are less than 0.1 for all periods, except for the HAR model.
From this result, we can conclude that our proposed DR-MVP model performs significantly better than the other models.

\begin{table}[!ht]
\caption{The $p$-values of the DM tests based on 10-minute portfolio risks, relative risks, and $\ell_2$-norm for the DRMVP-HAR model versus one of the DRMVP-50, HAR, FIVAR, and CLIME models for Period 1, Period 2, and the whole period.}
\label{Table-dm-relrisk}
\centering
\begin{tabular}{lrrrrlrrrr}
\hline
       & \multicolumn{4}{c}{Risk}                                                                                       &  & \multicolumn{4}{c}{Relative risk}                                                                               \\ \cline{2-5} \cline{7-10}
Period & \multicolumn{1}{c}{DRMVP-50} & \multicolumn{1}{c}{HAR} & \multicolumn{1}{c}{FIVAR} & \multicolumn{1}{c}{CLIME} &  & \multicolumn{1}{c}{DRMVP-50} & \multicolumn{1}{c}{HAR} & \multicolumn{1}{c}{FIVAR} & \multicolumn{1}{c}{CLIME}  \\ \hline
1      & 0.086                        & 1.000                   & 0.000                     & 0.014                     &  & 0.075                        & 1.000                   & 0.000                     & 0.007                      \\
2      & 0.000                        & 1.000                   & 0.000                     & 0.004                     &  & 0.000                        & 1.000                   & 0.000                     & 0.000                      \\
Whole  & 0.000                        & 1.000                   & 0.000                     & 0.000                     &  & 0.000                        & 1.000                   & 0.000                     & 0.000                      \\ \hline
       & \multicolumn{4}{c}{$\ell_2$-norm}                                                                                       &  & \multicolumn{4}{c}{}                                                                               \\ \cline{2-5} 
Period & \multicolumn{1}{c}{DRMVP-50} & \multicolumn{1}{c}{HAR} & \multicolumn{1}{c}{FIVAR} & \multicolumn{1}{c}{CLIME} &  &  &  &  &  \\ \cline{1-5} 
1      & 0.086                        & 1.000                   & 0.000                     & 0.014                     &  &  &  &  &  \\
2      & 0.000                        & 1.000                   & 0.000                     & 0.004                     &  &  &  &  &  \\
Whole  & 0.000                        & 1.000                   & 0.000                     & 0.000                     &  &  &  &  &  \\ \cline{1-5} 
\end{tabular}
\end{table}
To check the economic benefit of predicting MVPs, we calculated the Sharpe ratio as $\bar{ER} / S$,
where $\bar{ER}$ and $S$ are the mean and standard deviation of the excess return ${ER}_{i}$, respectively, ${ER}_{i} = \hat{\bar{\bw}}_{i}^{\top} R_{i} - r_{f,i}$, $R_{i}$ are close-to-close log-return vectors of $p$-stocks for day $i$, and $r_{f,i}$ is 3-month T-bill rate for day $i$.
Table \ref{Table-Sharpe} reports the out-of-sample Sharpe ratios of the normalized MVP with the DRMVP-50, DRMVP-HAR, HAR, FIVAR, and CLIME models for Period 1, Period 2, and the whole period.
Table \ref{Table-Sharpe} shows that the proposed DRMVP model does not significantly outperform the other models.
This may be because the primary objective of the DR-MVP model is to forecast MVP, which does not necessarily correspond to the point that maximizes the Sharpe ratio on the efficient frontier.
It would be an interesting future study to investigate and develop a model to capture the dynamics of the tangency portfolio that maximizes the Sharpe ratio.
We leave this for a future study.
\begin{table}[!ht]
\caption{Out-of-sample Sharpe ratios of the normalized MVP with the DRMVP-50, DRMVP-HAR, HAR, FIVAR, and CLIME models for Period 1, Period 2, and the whole period.}
\label{Table-Sharpe}
\centering
\begin{tabular}{lrrrrr}
\hline
Period & \multicolumn{1}{c}{DRMVP-50} & \multicolumn{1}{c}{DRMVP-HAR} & \multicolumn{1}{c}{HAR} & \multicolumn{1}{c}{FIVAR} & \multicolumn{1}{c}{CLIME}  \\ \hline
1      & 0.149                        & 0.155                         & 0.213                   & 0.177                     & 0.231                      \\
2      & 0.152                        & 0.142                         & 0.146                   & 0.086                     & 0.182                      \\
Whole  & 0.151                        & 0.148                         & 0.179                   & 0.133                     & 0.206                      \\ \hline
\end{tabular}
\end{table}

\section{Conclusion}\label{SEC-6}

In this paper, we proposed a novel dynamic realized MVP model that can accommodate nonlinear volatility dynamics by directly explaining the MVP dynamics. 
To obtain non-parametric realized MVPs for each day, we employed the CLIME estimation procedure under the sparse condition for the inverse volatility matrix.
With the non-parametric realized MVP, we found the AR dynamic structure for the realized MVP and constructed dynamic models to explain the AR structure.
To connect the high-frequency-based information and low-frequency dynamics, we suggested a diffusion process that would provide the rigorous mathematical background.
The empirical study shows that modeling the dynamic structure of the realized MVP helps account for the market dynamics in terms of the minimum variance portfolio.

\section*{Acknowledgments} 
 
The research of Donggyu Kim was supported in part by the National Research Foundation of Korea (NRF) (2021R1C1C1003216).

\bibliography{myReferences}

\clearpage

\appendix
\setcounter{figure}{0}
\renewcommand{\thefigure}{A\arabic{figure}}
\section{Volatility matrix process for the DR-MVP model}\label{SEC:sample-model}

\subsection{General DR-MVP model} \label{cond-general}

The proposed DR-MVP model originated from the inverse matrix dynamic model in \eqref{eq:inversematrixmodel}.
In order to provide an example volatility matrix process that satisfies the DR-MVP model, it is beneficial to use the BEKK($0,q$) structure inverse matrix dynamics since BEKK($0,q$) structure inherently ensures the positive semidefiniteness itself.
In Section 2, we provided some conditions on $\bbeta$ and $\bOmega$ satisfying that if the inverse integrated volatility matrix follows the inverse matrix dynamic of BEKK($0,q$) form, then the MVP has the VAR($q$) structure.
Although the conditions do not hold, we can provide a volatility matrix process satisfying given MVP dynamics of VAR($q$) in \eqref{eq:g-AR}, using the inverse matrix dynamics of BEKK($0,q$) in \eqref{eq:G-AR}.
Suppose that the integrated volatility matrix consists of two parts, where one of its inverse matrices follows the inverse matrix dynamics of BEKK($0,q$), while the difference between the inverse of the integrated volatility matrix and that of the aforementioned matrix remains diagonal.
Specifically, we have
\begin{align*}
  & \bGamma_d = \bGamma_{1,d} + \bGamma_{2,d} , \quad \mathbb{E}\left[ \bGamma_{1,d}^{-1} \big| \mathcal{F}_{d-1} \right] = \bB_{0} \bB_{0}^{\top} + \sum_{j=1}^{q} \bB_{j} \bGamma_{1,d-j}^{-1} \bB_{j}^{\top} , \quad \text{and} \quad \bOmega_d - \bGamma_{1,d}^{-1} \text{ is diagonal,}
\end{align*}
where $\bGamma_d = \int_{d-1}^{d} \bSigma_{t} dt$, $\bGamma_{1,d} = \int_{d-1}^{d} \bSigma_{1,t} dt$, $\bGamma_{2,d} = \int_{d-1}^{d} \bSigma_{2,t} dt$, and $\bSigma_t=\bSigma_{1,t} + \bSigma_{2,t}$.
Using Woodbury matrix identity, we have
\begin{eqnarray}\label{eq:volpart}
  \bOmega_{d} &=&  (\bGamma_{1,d} + \bGamma_{2,d})^{-1} \cr
  &=& \bGamma_{1,d}^{-1} - (\bGamma_{1,d} + \bGamma_{1,d}\bGamma_{2,d}^{-1}\bGamma_{1,d})^{-1} \cr
  &=& \bGamma_{1,d}^{-1} - \tilde{\bGamma}_{2,d}^{-1}
  ,
\end{eqnarray}
where $\tilde{\bGamma}_{2,d} = \bGamma_{1,d} + \bGamma_{1,d}\bGamma_{2,d}^{-1}\bGamma_{1,d}$.
To satisfy $\mathbb{E}\left[ \bOmega_d \b1 \big| \mathcal{F}_{d-1} \right]  = \bbeta_0 + \sum_{j=1}^{q} \bbeta_{j} \bw_{d-j}$, it is required to satisfy the following equation:
\begin{equation*}
  \br_{d} \coloneqq \mathbb{E}\left[ \tilde{\bGamma}_{2,d}^{-1} \b1 \big| \mathcal{F}_{d-1} \right] = \bB_{0} \bB_{0}^{\top} \b1 + \sum_{j=1}^{q} \bB_{j} \bGamma_{1,d-j}^{-1} \bB_{j}^{\top} \b1 - \left( \bbeta_0 + \sum_{j=1}^{q} \bbeta_{j} \bw_{d-j} \right) 
  .
\end{equation*}
We note that the above equation indicates the inverse matrix dynamics for $\tilde{\bGamma}_{2,d}$.
Simple algebra shows that
\begin{eqnarray*}
  && \br_{d,i} - \left( \bB_{0} \bB_{0}^{\top} \b1 - \bbeta_{0}  \right)_i \cr
  &=& \sum_{s=1}^{q} \left( \sum_{j,k,l=1}^{p} \bB_{s,ij} \bGamma_{1,d-s,jk}^{-1} \bB_{s,lk}^{\top} - \sum_{j,k=1}^{p} \bbeta_{s,ij} \bOmega_{d-s,jk} \right) \cr
  &=& \sum_{s=1}^{q} \left( \sum_{j,k} \bB_{s,\cdot k} \bB_{s,ij} \bGamma_{1,d-s,jk}^{-1} - \sum_{j,k} \bbeta_{s,ij} \bOmega_{d-s,jk} \right) \cr
  &=& \sum_{s=1}^{q} \Bigg( \sum_{j,k} \bB_{s,\cdot k} \bB_{s,ij} \bOmega_{d-s,jk} - \sum_{j,k} \bbeta_{s,ij} \bOmega_{d-s,jk}  + \sum_{j,k} \bB_{s,\cdot k} \bB_{s,ij} \tilde{\bGamma}_{2,d-s,jk}^{-1} \Bigg)  \cr
  &=& \sum_{s=1}^{q} \Bigg( \sum_{j,k} \bB_{s,\cdot k} \bB_{s,ij} \bOmega_{d-s,jk} - \sum_{j,k} \bbeta_{s,ij} \bOmega_{d-s,jk}  \Bigg)  + \sum_{s=1}^{q} \sum_{j}^{p} \bB_{s,\cdot j} \bB_{s,ij} \br_{d-s,j}
  ,
\end{eqnarray*}
where the last equality is due to tower property.
Therefore, $\br_d$ follows the VAR model with additional innovation terms.
For given $\bbeta_0, \ldots, \bbeta_{q}$, we can choose $\bB_{0}, \ldots, \bB_{q}$ that make $\br_{d}$ negative.
Then, $\bOmega_{d} = \bGamma_{1,d}^{-1} - \tilde{\bGamma}_{2,d}^{-1}$ is positive semidefinite because $\bGamma_{1,d}^{-1}$ is positive semidefinite and $\tilde{\bGamma}_{2,d}^{-1}$ is negative diagonal matrix.
That is, there exist inverse matrix dynamics satisfying the given DR-MVP model.

\subsection{Positive semidefiniteness of the instantaneous volatility process}\label{cond-pd}
To satisfy the positive semidefiniteness of the instantaneous volatility process in \eqref{eq:DRMVPvol}, we need some conditions.
We provide the sufficient condition as follows:
\begin{enumerate}
  \item For any $d \in \mathbb{N}$ and $t \in (d-1, d]$, $\bSigma_{d-1}$, $\bG_{d}(\btheta)$, and $\bM_{t}$ have the same eigenvectors $\bU$.
  \item For any $1 \leq j \leq p$, $d \in \mathbb{N}$, and $t \in (d-1, d]$, we have $m_{t,j} < \sigma_{d-1,j}^{-1} - 2 \langle t \rangle (\sigma_{d-1,j}^{-1} - g_{d,j})$, where $g_{d,j} = [\bU^{\top}\bG_{d}(\btheta)\bU]_{jj}$, $\sigma_{d-1,j} = [\bU^{\top}\bSigma_{d-1}\bU]_{jj}$, and $m_{t,j} = [\bU^{\top}\bM_{t}\bU]_{jj}$.
\end{enumerate}
The first condition is related to time-invariant factor loading, which is often assumed to model dynamic high-dimensional volatility matrix processes \citep{kim2020unified,kim2023factor,shin2021factor}.
The second condition bounds the maximum random fluctuation of the inverse integrated volatility matrix to ensure that the spot volatility matrix remains positive semidefinite.

\subsection{Data generating process}
To generate a volatility matrix process that satisfies the given DR-MVP model in \eqref{eq:g-AR}, we utilize the idea of partitioning the integrated volatility matrix in Section \ref{cond-general}.
We assume that the log-price of assets follow
\begin{equation*}
  d\bX_{t} = \bmu_{t} dt + (\bSigma_{1,t} + \bSigma_{2,t})^{1/2} d{\mathcal{B}}_{t}
  ,
\end{equation*}
where $\bmu_{t}$ is a drift process, ${\mathcal{B}}_{t}$ is a $p$-dimensional standard Brownian motion, $\bSigma_{1,t}$ and $\bSigma_{2,t}$ follow the notations in the example of partitioning the integrated volatility matrix.
Let $\bGamma_{1,t} = \int_{[t-]}^{t} \bSigma_{1,t} dt$, $\bGamma_{2,t} = \int_{[t-]}^{t} \bSigma_{2,t} dt$, and $\tilde{\bGamma}_{2,t} = \bGamma_{1,t} + \bGamma_{1,t}\bGamma_{2,t}^{-1}\bGamma_{1,t}$ is diagonal matrix process.
We suppose that $\bSigma_{1,t}$ follow \eqref{eq:DRMVPvol} with $\bG_{1,d} = \bB_{0} \bB_{0}^{\top} \b1 + \sum_{j=1}^{q} \bB_{j} \bGamma_{1,d-j}^{-1} \bB_{j}^{\top} \b1$.
Then, by Proposition 1, we have $\mathbb{E}\left[ \bGamma_{1,d}^{-1} \big| \mathcal{F}_{d-1} \right] = \bG_{1,d}$.
Simple algebra shows that
\begin{equation*}
  \bGamma_{2,t} = \bGamma_{1,t} (\tilde{\bGamma}_{2,t} - \bGamma_{1,t})^{-1} \bGamma_{1,t}
  .
\end{equation*}
By differentiating both sides, we have
\begin{equation}\label{eq:generate-sigma2}
  \bSigma_{2,t} = \bSigma_{1,t} \bP_{t} + \bP_{t}^{\top} \bSigma_{1,t} + \bP_{t}^{\top} \left( \Pi_{t} - \bSigma_{1,t} \right) \bP_{t}
  ,
\end{equation}
where $\bP_{t} = (\tilde{\bGamma}_{2,t} - \bGamma_{1,t})^{-1} \bGamma_{1,t}$ and $\Pi_{t} = \frac{d}{dt} \tilde{\bGamma}_{2,t}$.
Since $\tilde{\bGamma}_{2,t}$ is diagonal, $\Pi_t$ is also diagonal.
We can generate $\Pi_{t}$ instead of directly generating $\bSigma_{2,t}$.
If we generate the diagonal matrix process $\Pi_t$ that follows \eqref{eq:DRMVPvol} with the conditional inverse integrated matrix
\begin{equation*}
  \mathbb{E}\left[ \left( \int_{[d-1]}^{d} \Pi_t dt \right) ^{-1} \b1  \big| \mathcal{F}_{d-1} \right] = \bB_{0} \bB_{0}^{\top} \b1 + \sum_{j=1}^{q} \bB_{j} \bGamma_{1,d-j}^{-1} \bB_{j}^{\top} \b1 - \left( \bbeta_0 + \sum_{j=1}^{q} \bbeta_{j} \bw_{d-j} \right) 
  ,
\end{equation*}
then the MVP weights follow the DR-MVP model of AR($q$) structure,
\begin{equation*}
  \mathbb{E}\left[ \bOmega_d \b1 \big| \mathcal{F}_{d-1} \right] = \bbeta_{0} + \sum_{j=1}^{q} \bbeta_j \bw_{d-j}
  .
\end{equation*}

In the simulation study, we consider the HAR-type DR-MVP model
$$\bg_d(\btheta) = \bbeta_0 + \sum_{j\in \left\lbrace 1,5,22 \right\rbrace} \bbeta_{j} {\bw}_{d}^{(j)} ,$$
where ${\bw}_{d}^{(j)}=j^{-1} \sum_{k=1}^{j} \bw_{d-k}$.
To generate a volatility process that satisfies the DR-MVP model, we specifically follow the subsequent data generating process:
\begin{enumerate}
  \item Set the coefficients $\bbeta_{1} = 0.3 \bbeta$, $\bbeta_{5} = 0.6 \bbeta$, $\bbeta_{22} = 0.1 \bbeta$, where
  \begin{equation*}
  [\bbeta]_{ij} = \begin{cases}
    0.2, & \text{if } i=j,\cr
    0.7, & \text{if } i=j+1,\cr
    0, & \text{otherwise. }
  \end{cases}
\end{equation*}
  \item Set $\bbeta_0 = \bU \diag(\bq) \bU^{\top} \b1 + q_0 \b1$, where $\bU$ are eigenvectors of $\bbeta+\bbeta^{\top}$, $\bq_{1}=40$, $\bq_{2}=80$, $ \bq_{3}=120$, $\bq_{i}=246+i$ for $4 \leq i \leq 100$, and $q_0=800$.
  We then set $B_0=\bU \diag(\sqrt{\bq}) \bU^{\top}$ and $[\bB_{j}]_{kl} = (\bar{\bbeta}_{j,kl}) / \sqrt{\sum_{r=1}^{p} \bar{\bbeta}_{j,rl}}$ for $j\in \left\lbrace 1,5,22 \right\rbrace$, where $\bar{\bbeta}_{j} = (\bbeta_{j} + \bbeta_{j}^{\top})/2$.
  The choice of $\bq$ is motivated by the factor model for the volatility matrix.
  \item Generate $\bSigma_{1,t}$ that follows \eqref{eq:DRMVPvol}, where the conditional expectation of its inverse integrated matrix is
  $\bG_{1,d} = \bB_{0} \bB_{0}^{\top} + \sum_{j \in \left\lbrace 1,5,22 \right\rbrace} \bB_{j} {\bOmega}_{d}^{(j)} \bB_{j}^{\top}$, ${\bOmega}_{d}^{(j)} = j^{-1} \sum_{k=1}^{j} \bGamma_{d-k}^{-1}$,
  $\bM_{1,t} = \bU \diag(\bm_{1,t}) \bU^{\top}$, $\bm_{1,k,t} = \min(\max(u_{1,k,d-1}\times(Z_{1,k,t}-Z_{1,k,d-1}),-0.2u_{1,k,d-1}),0.2u_{1,k,d-1})$,
  $(Z_{1,k,t})_{k=1,\ldots,p}$ are independent standard Brownian motions,
  and $u_{1,k,d-1} = [\bU^{\top} \bSigma_{1,d-1} \bU]_{kk}$ for $t \in (d-1,d]$.
  \item Generate a diagonal matrix process $\Pi_{t}$ that follows \eqref{eq:DRMVPvol}, where the conditional expectation of its inverse integrated matrix is
  $\diag \left(  \bG_{1,d} \b1 - \bbeta_0 - \sum_{j\in \left\lbrace 1,5,22 \right\rbrace} \bbeta_{j} {\bw}_{d}^{(j)}   \right),$ $\bM_{1,t} = \bU \diag(\bm_{1,t}) \bU^{\top}$, $\bm_{1,k,t} = \min(\max(u_{2,k,d-1}\times (Z_{2,k,t} - Z_{2,k,d-1}) , -u_{2,k,d-1}),$ $u_{2,k,d-1})$, $(Z_{2,k,t})_{k=1,\ldots,p}$ are independent standard Brownian motions, and $u_{2,k,d-1} = [ \Pi_{d-1} ]_{kk}$ for $t \in (d-1,d]$.
  We set the initial value of the process $\Pi_{t}$ as $\Pi_{0} = \diag \Big(   \bG_{1,d} \b1 - \bbeta_0 - \sum_{j\in \left\lbrace 1,5,22 \right\rbrace} \bbeta_{j} {\bw}_{d}^{(j)}   \Big) ^{-1}$, the inverse of the conditional expectation of its inverse integrated matrix.
  From $\bSigma_{1,t}$ and $\Pi_t$, we calculate $\bSigma_{2,t}$ using \eqref{eq:generate-sigma2}.
  \item Generate the true log-prices of assets as $d\bX_{t}=(\bSigma_{1,t}+\bSigma_{2,t})^{1/2} d{\mathcal{B}}_{t} + L_{t} d\bLambda_{t}$, where the jump sizes $L_{t}$'s obey the normal distribution with mean $0.05$ and variance $0.005^2$, the signs of $L_t$ are randomly determined, and $\bLambda_{t}$ is standard Poisson process with intensity $5$.
  We then generate the observed log-prices of assets as $\bY_{t_{d,i}} = \bX_{t_{d,i}} + \be_{t_{d,i}}$, where the noises $\be_{t_{d,i}}$'s obey the independent normal distributions with mean zero and variances $0.01 \times \diag(\bSigma_{1,t_{d,0}}+\bSigma_{2,t_{d,0}})$.
\end{enumerate}

\section{JPRVM estimator}\label{SEC:jprvm}
In Sections \ref{SEC-4} and \ref{SEC-5}, in order to estimate the integrated volatility matrix, we employed the JPRVM estimator \citet{ait2016increased} as follows:
\begin{equation}\label{eq:SimulationPRVM}
	\[\hat{\bGamma}_{d}^{JPRVM}\]_{ij} = \frac{1}{w \phi} \sum_{u=1}^{m - w + 1} \left (\bar{X}_{d,u}^{i} \bar{X}_{d,u}^{j} - \frac{1}{2} \hat{X}_{d,u}^{ij}  \right ) \mathbbm{1}\left(\abs{\bar{X}_{d,u}^{i} < v_d^{i} }\right) \mathbbm{1}\left(\abs{\bar{X}_{d,u}^{j} < v_d^{j} }\right),
\end{equation}
where $w = [m^{1/2}]$ is a pre-averaging window suggested in \citet{christensen2010pre}, $\phi = 1/12$, $g(x)=x \land (1-x)$,
$\hat{X}_{d,u}^{ij}= \sum_{s=1}^{w} (g(\frac{s}{w}) - g(\frac{s-1}{w}))^2 \Delta^{d}_{u+s-1} X_{i} \Delta^{d}_{u+s-1} X_{j}$,
$\Delta^{d}_{u}X_{i} = \bX_{i,t_{d-1,u}} - \bX_{i,t_{d-1,u-1}}$,
$\bar{X}_{d,u}^{i}= \sum_{s=1}^{w-1} g(s/w) \Delta^{d}_{u+s} X_{i}$, and
$v_d^{i} = 3(w/m)^{0.47} \sqrt{w^{-1} \sum_{u=1}^{m-w+1} ( \bar{X}_{d,u}^{i} )^2 } $ is a jump truncation parameter, as suggested in \citet{ait2016increased}.
\section{Additional simulation analyses}\label{SEC:appendix-simulation}

\begin{figure}[!h]
\centering
\includegraphics[width = 1\textwidth]{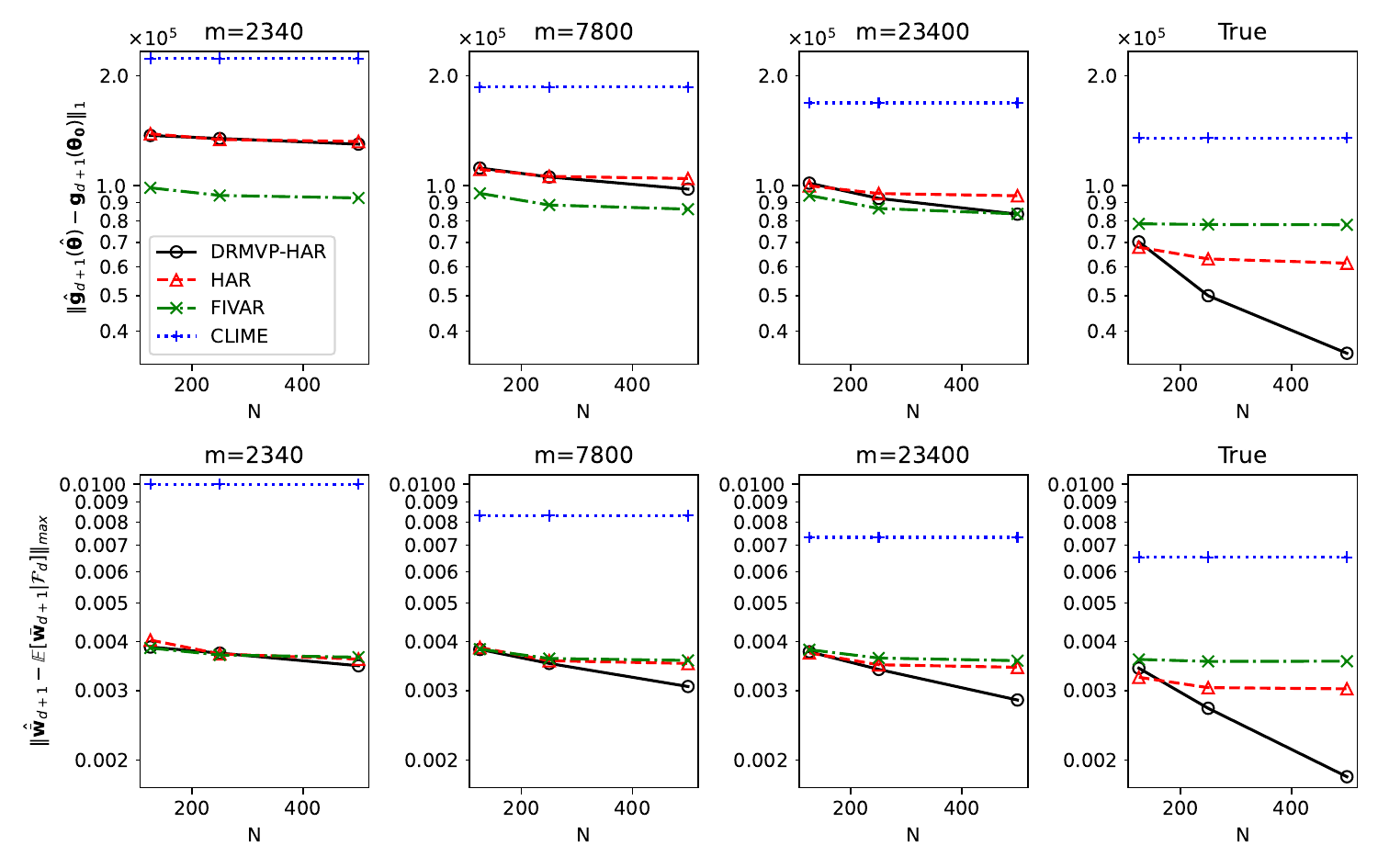}
\caption{Mean of $\lVert \hat{\bg}_{d+1}(\hat{\btheta}) - \bg_{d+1} (\btheta_{0}) \rVert_{1}$ and $\lVert \hat{\bar{\bw}}_{d+1} - \mathbb{E}\left[ \bar{\bw}_{d+1} | \mathcal{F}_{d} \right] \rVert_{\max}$ for $N=125,250,500$ and $m=2340,7800,23400$, along with cases employing the true values in place of nonparametric estimators.}
\label{fig:PredictSimulationAppendix}
\end{figure}
We further checked the mean of $\lVert \hat{\bg}_{d+1}(\hat{\btheta}) - \bg_{d+1} (\btheta_{0}) \rVert_{1}$ and $\lVert \hat{\bar{\bw}}_{d+1} - \mathbb{E}\left[ \bar{\bw}_{d+1} | \mathcal{F}_{d} \right] \rVert_{\max}$.
Figure \ref{fig:PredictSimulationAppendix} draws the mean of $\lVert \hat{\bg}_{d+1}(\hat{\btheta}) - \bg_{d+1} (\btheta_{0}) \rVert_{\max}$ and $\lVert \hat{\bar{\bw}}_{d+1} - \mathbb{E}\left[ \bar{\bw}_{d+1} | \mathcal{F}_{d} \right] \rVert_{1}$ for $N=125,250,500$ and $m=2340,7800,23400$, along with cases employing the true values in place of nonparametric estimators.
From Figure \ref{fig:PredictSimulationAppendix}, we find that the mean prediction errors of the conditional non-normalized and normalized MVPs decrease as the number of high- or low-frequency observations increases.
In the case of $\lVert \hat{\bg}_{d+1}(\hat{\btheta}) - \bg_{d+1} (\btheta_{0}) \rVert_{1}$, the models using CLIME estimators perform worse than the FIVAR model.
This may be because of the bias of the CLIME estimator.
Due to the bias in the CLIME estimator, the $\ell_1$-norm error of $\hat{\bg}_{d+1}(\hat{\btheta})$, $\lVert \hat{\bg}_{d+1}(\hat{\btheta}) - \bg_{d+1} (\btheta_{0}) \rVert_{1}$, can have the order $p \times s_{\beta} \lambda_{n}$.
We can find that when they employ the true values instead of nonparametric estimations, the DR-MVP models perform better than the FIVAR model since there are no biases in the input MVPs.
In the case of $\lVert \hat{\bar{\bw}}_{d+1} - \mathbb{E}\left[ \bar{\bw}_{d+1} | \mathcal{F}_{d} \right] \rVert_{\max}$, the DRMVP-HAR and HAR model performs better than the other models as the number of high- and low-frequency observations increase.

To assess the robustness of the proposed methodology when the true model is misspecified, we conducted an additional analysis assuming that the data-generating process follows a different specification, namely the FIVAR model \citep{shin2021factor}.
We used the same data-generating process as that of Section 5.1 in \citet{shin2021factor}, except for the number of assets, $p=100$ instead of $200$.
\begin{figure}[!h] 
	\centering
	\includegraphics[width = 0.85\textwidth]{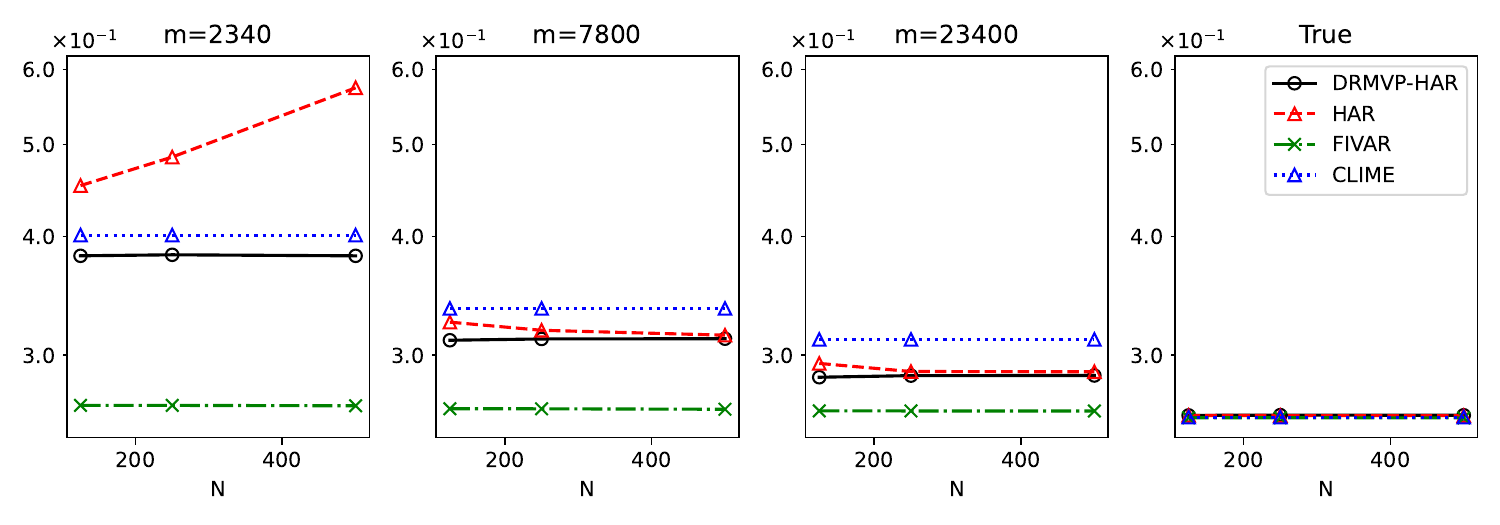} 
	\caption{Mean out-of-sample of normalized portfolio risk for the DR-MVP, FIVAR, and CLIME models with $N=125,250,500$ and $m=2340,7800,23400$, along with cases employing the true values in place of nonparametric estimators, under the FIVAR data-generating process.}
	\label{fig:MisspecifiedRisk}
\end{figure}

Figure \ref{fig:MisspecifiedRisk} draws the mean out-of-sample normalized portfolio risk for the DR-MVP, HAR, FIVAR, and CLIME models under the FIVAR data-generating process.
From Figure \ref{fig:MisspecifiedRisk}, we find that the performances of proposed DR-MVP methods get closer to the performance of the FIVAR model as $m$ increases, while the effect of increasing $n$ is insignificant.
There may be two reasons why the proposed DR-MVP models show comparable performances under the misspecified model.
First, mathematically, Taylor's expansion implies that the  FIVAR dynamics can be partially explained by the linear dynamics model of its inverse, such as the proposed DR-MVP model. 
Second, the target of the DR-MVP model is to predict the non-normalized minimum variance portfolio, while that of the FIVAR model is to predict the whole volatility matrix.
Thus, to estimate the MVP weights, the FIVAR procedure needs to estimate the inverse matrix of the predicted volatility matrix. 
This may cause some estimation errors. 
In fact, in the case of employing the true values instead of the nonparametric estimations, the FIVAR model does not show the best performance.
This may be because although the FIVAR model uses the true integrated volatility matrices as inputs to obtain $\hat{\bGamma}^{\text{F}}_{d+1}$, there exists an estimation error to obtain the inverse of $\hat{\bGamma}^{\text{F}}_{d+1}$.
At the same time, there is no estimation error in obtaining inverse matrices for the DR-MVP, HAR, and CLIME models since they employ the true MVPs.
On the other hand, incorporating higher frequency data helps estimate MVP, while utilizing more in-samples (low-frequency data) is less beneficial due to the model misspecification.

  \section{Comparison between integrated volatility and quadratic covariation }\label{compare-IQ}

 To check the effect of the jumps, we additionally conducted the empirical analysis of out-of-sample annualized risks for the proposed procedure with quadratic covariations.
 Table \ref{Table-IQ} reports the out-of-sample annualized risks of the normalized MVP with the DRMVP-50, DRMVP-HAR, HAR, and CLIME models.
 We note that the risk is measured by the 10-minute portfolio log-returns including jumps. 
From Table \ref{Table-IQ}, we find that the DR-MVP models using the integrated volatility perform better than the DR-MVP models using the quadratic variation.
This may be because the jump components are highly unpredictable.
 
\begin{table}[!ht]
\caption{Out-of-sample annualized risks of the normalized MVP for integrated volatility (I) and quadratic covariation (Q) with the DR-MVP, HAR, FIVAR, and CLIME models for Period 1, Period 2, and the whole period.}
\label{Table-IQ}
\centering
\scalebox{1}{
\begin{tabular}{lrrlrrlrrlrr}
\hline
       & \multicolumn{2}{c}{DRMVP-50}                  &  & \multicolumn{2}{c}{DRMVP-HAR}                 &  & \multicolumn{2}{c}{HAR}                       &  & \multicolumn{2}{c}{CLIME}                      \\ \cline{2-3} \cline{5-6} \cline{8-9} \cline{11-12}
Period & \multicolumn{1}{c}{I} & \multicolumn{1}{c}{Q} &  & \multicolumn{1}{c}{I} & \multicolumn{1}{c}{Q} &  & \multicolumn{1}{c}{I} & \multicolumn{1}{c}{Q} &  & \multicolumn{1}{c}{I} & \multicolumn{1}{c}{Q}  \\ \hline
1      & 4.667\%               & 5.548\%               &  & 4.660\%               & 5.486\%               &  & 4.512\%               & 5.329\%               &  & 4.801\%               & 5.295\%                \\
2      & 4.199\%               & 14.385\%              &  & 4.182\%               & 4.564\%               &  & 4.124\%               & 4.846\%               &  & 4.333\%               & 4.510\%                \\
Whole  & 4.438\%               & 10.918\%              &  & 4.427\%               & 5.044\%               &  & 4.322\%               & 5.092\%               &  & 4.572\%               & 4.917\%                \\ \hline
\end{tabular}
}
\end{table}

    \section{Proofs}\label{SEC-7}

  \subsection{Proof of Theorem \ref{prop:DRMVPvol}}

\textbf{Proof of Theorem \ref{prop:DRMVPvol}.}
Simple algebra shows that \eqref{eq:DRMVPvol} is equivalent to
\begin{equation}\label{eq:DRMVPvolAA}
   - \bSigma_{d-1}^{-1}  - (4 \langle t \rangle  - 3 \langle t \rangle ^{2})  (\bSigma_{d-1}^{-1} - \bG_{d}(\btheta))  +  \bM_{t} = - \bA_{t-}^{-1} \bSigma_{t} \bA_{t-}^{-1} + 2 \bA_{t-} ,
\end{equation}
By integrating both sides of \eqref{eq:DRMVPvolAA}, we have
\begin{eqnarray*}
  \bG_{d}(\btheta) + \int_{d-1}^{d} \bM_{s} ds  
  &=& \int_{d-1}^{d} (- \bA_{t-}^{-1} \bSigma_{t} \bA_{t-}^{-1} + 2 \bA_{t-})  dt \cr
  &=& \int_{d-1}^{d}  (\bA_{t-}^{-1} - \bA_{t-}^{-1} \bSigma_{t} \bA_{t-}^{-1} ) dt + \bA_{t-}^{-1} dt \cr
  &=& \int_{d-1}^{d}   \langle t \rangle d(\bA_{t}^{-1} ) + \bA_{t}^{-1} dt \cr
  &=& \int_{d-1}^{d} d(\langle t \rangle \bA_{t}^{-1} ) \cr
  &=& \left( \int_{d-1}^{d} \bSigma_{t} dt \right) ^{-1}
  .
\end{eqnarray*}
Therefore, we obtain $\bOmega_{d} = \bG_{d}(\btheta) + \bU_{d}$ for any $d \in \mathbb{N}$, where $\bU_{d} = \int_{d-1}^{d} \bM_{t} dt$.
$\blacksquare$

  \subsection{Proof of Proposition \ref{prop1}}
  
  \textbf{Proof of Proposition \ref{prop1}.} 
We first consider \eqref{prop1-result1}. 
We have
\begin{eqnarray*}
	&&E \[ \bar{\bw}_d \middle | \FF_{d-1} \] - 	\frac{\bg_{d} (\btheta)}{\b1^{\top} \bg_{d} (\btheta)} \cr
	&&=  E \[ \frac{\bg_{d} (\btheta)}{\b1^{\top} \bg_{d} (\btheta) + \b1 ^{\top} \bepsilon_{d} } -  \frac{\bg_{d} (\btheta)}{\b1^{\top} \bg_{d} (\btheta)}   \middle | \FF_{d-1}  \]  +  E \[ \frac{\bepsilon_{d}}{\b1^{\top} \bg_{d} (\btheta) + \b1 ^{\top} \bepsilon_{d} } \middle | \FF_{d-1}   \] \cr
	&&= (I) + (II).
\end{eqnarray*}
For $(I)$, we have 
\begin{eqnarray} \label{prop1-eq-1}
	 \| (I) \|_{\max}  &\leq&    \left \| \frac{\bg_{d} (\btheta)}{\b1^{\top} \bg_{d} (\btheta)}  \right \|_{\max}  E \[ \left | \frac{  p^{-1} \b1 ^{\top} \bepsilon_{d} }{   p^{-1} \b1^{\top} \bg_{d} (\btheta) +   p^{-1} \b1 ^{\top} \bepsilon_{d} } \right | \middle |      \FF_{d-1} \] \cr
	 	&\leq& \left \| \frac{\bg_{d} (\btheta)}{\b1^{\top} \bg_{d} (\btheta)}  \right \|_{\max}  E \[ \left |    p^{-1} \b1 ^{\top} \bepsilon_{d}  \right | ^2  \middle |      \FF_{d-1} \] ^{1/2}  E \[ \left | \frac{  1}{   p^{-1}  \b1 ^{\top} \bOmega_{d} \b1  } \right | ^2 \middle |      \FF_{d-1} \] ^{1/2}  \cr
	 	&\leq& C   \left \| \frac{\bg_{d} (\btheta)}{\b1^{\top} \bg_{d} (\btheta)}  \right \|_{\max}  p^{-1/2} \text { a.s.},
\end{eqnarray}
where the last inequality is due to Assumption \ref{assumption1}.
For $(II)$, we have
\begin{eqnarray*}
 	\| (II) \|_{\max} &=&  \left \| E \[ \frac{\bepsilon_{d}}{\b1^{\top} \bg_{d} (\btheta) + \b1 ^{\top} \bepsilon_{d} }  - \frac{\bepsilon_{d}}{\b1^{\top} \bg_{d} (\btheta)   } \middle | \FF_{d-1}   \] \right \|_{\max} \cr
 	&\leq&  \left \| E \[ \frac{\bepsilon_{d}}{\b1^{\top} \bg_{d} (\btheta)   }   \left \{  \frac{\b1 ^{\top} \bepsilon_{d} }{\b1^{\top} \bg_{d} (\btheta) + \b1 ^{\top} \bepsilon_{d} }    \right \} \middle | \FF_{d-1}   \] \right \|_{\max} \cr
 	 	&\leq&   E \[  \left \| \frac{\bepsilon_{d}}{\b1^{\top} \bg_{d} (\btheta)   }   \right \|_{\max} ^2  \middle | \FF_{d-1} \] ^{1/2} E\[  \left |  \frac{\b1 ^{\top} \bepsilon_{d} }{\b1^{\top} \bg_{d} (\btheta) + \b1 ^{\top} \bepsilon_{d} }    \right |^2 \middle | \FF_{d-1}   \] ^{1/2} \cr
 	 	&\leq&  C p^{-3/2} E \[  \left \|  \bepsilon_{d}  \right \|_{\max} ^2  \middle | \FF_{d-1} \] ^{1/2} \cr
    &\leq& C p^{-3/2} \sqrt{\log p} \text{ a.s.},
\end{eqnarray*}
where the third inequality can be shown similar to the proof of \eqref{prop1-eq-1} and the last inequality is due to the sub-Gaussianity of $\epsilon_{i,d}$'s. 
Thus, the statement \eqref{prop1-result1} is proofed. 
Similarly, we can show \eqref{prop1-result2} using the fact that the smallest value of  $  \left \|  	\frac{\bg_{d} (\btheta)}{\b1^{\top} \bg_{d} (\btheta)}  \right  \|_{1}$ is one.
\endpf

 \subsection{Proof of Proposition \ref{thm1}}
 
 \textbf{Proof of Proposition \ref{thm1}.}
 Under Assumption \ref{assumption2}(a), we have, for some large constant  $C_c$.
  \begin{equation*}
 	P \( \max_d \max_{i,j} |\hat{\Gamma}_{ij, d} - \Gamma_{ij,d} | \geq C_c \sqrt{\log (p \vee N) } m^{-1/4}   \) \leq C (p \vee N )^c. 
 \end{equation*}
 Thus, we show the statements based on the event:
\begin{equation}\label{thm1-eq1}
 	\max_{d \leq N} \| (\hat{\bGamma}_d -\bGamma_d ) \|_{\max} \leq   C   \sqrt{\log (p \vee N) } m^{-1/4}.
\end{equation}

 Consider \eqref{thm-result1}. 
 We have
 \begin{eqnarray*}
 	\max_{d \leq N} \| \hat{\bGamma}_d \Omega_d - \bI \|_{\max} &\leq&   \max_{d \leq N} \| (\hat{\bGamma}_d -\bGamma_d ) \|_{\max} \| \Omega_d  \|_{1}  \cr
 	&\leq& C   \sqrt{\log (p \vee N) } m^{-1/4},
 \end{eqnarray*}
 where the second inequality is due to \eqref{thm1-eq1}. 
 Thus, $\bOmega_d$ satisfies $\|\hat{\bGamma}_d   \bA - \bI  \|_{\max} \leq \tau_m$.
 Therefore, we have
 \begin{equation*}
 	\max_{d} \|\hat{\bOmega}_d\|_{1}  \leq  	\max_{d} \|\bOmega_d\|_{1}. 
 \end{equation*}
 Then, we have
 \begin{eqnarray*}
 	\max_{d \leq N} \| \hat{\bOmega}_d -\bOmega_d\|_{\max} &\leq&  \max_{d \leq N}  \|  \bOmega_d \|_1 \| \hat{\bGamma}_d  \hat{\bOmega}_d -\bI \|_{\max}  + \max_{d \leq N}  \|\bOmega_d  \hat{\bGamma}_d  - \bI \|_{\max}  \| \hat{\bOmega}_d\|_1 \cr
 		&\leq& C \tau_m.
 \end{eqnarray*}
 Similar to proofs of Theorem 6 in \citet{cai2011constrained}, we can show the $\ell_1$ bound.

  Consider \eqref{thm-result2}.
 We have 
 \begin{eqnarray*}
  \max_{j} |\hat{w}_{j,d} -w_{j,0d}|  &\leq&  \max_{j} \sum_{i=1}^p | \hat{\Omega}_{ij d} -  \Omega_{ij d} | \cr
  &=& \|  \hat{\bOmega}_d - \bOmega_d \| _1 \cr
  &\leq& C s_p \tau_m ^{1-\delta},
\end{eqnarray*}
 where the last inequality is due to \eqref{thm-result1}.
 \endpf

  \subsection{Proof of Theorem \ref{thm-2}}

 \textbf{Proof of Theorem \ref{thm-2}.} 
 We have
 \begin{eqnarray*}
	\max_i | \hat{g}_{i,d} (\btheta_{0i}) - g_{i,d} (\btheta_{0i})| &\leq&  \max_i \| \btheta_{0i} \|_1 \max_{j\leq d} \| \hat{\bw}_{j} -\bw_{j} \|_{\max} \cr
		&\leq& C s_{\beta} s_p \tau_m ^{1-\delta},
 \end{eqnarray*}
 where the last inequality is due to Assumption \ref{assumption3}(d).
 Similarly, we can show
  \begin{eqnarray*}
 	\max_i \| \frac{  \hat{g}_{i,d} (\btheta_{0,i}) }{\partial \btheta_i }\| _{\max} \leq  \max_d \| \hat{\bw}_d -\bw_d \|_{\max}  \leq   C  s_p \tau_m ^{1-\delta}.
  \end{eqnarray*}
 By Assumption \ref{assumption2}(b), we have
 \begin{equation*}
 	\| \bw_d \|_{\max}  = \| \bOmega_d \b1\|_{\max} \leq  \|\bOmega_d\|_1 \leq C \text{ a.s}.
 \end{equation*} 
 Thus, $\bw_d$'s are bounded random variables.  
 Then, we have 
 \begin{equation*}
 	\| \nabla \mathcal{L}_{n,i} (\btheta_{0,i}) \| _{\max}  \leq  \left \|  \frac{2 }{n} \sum_{d=1}^n \( w_{i,d} - g_{i,d} (\btheta_{0i}) \)  \frac{  g_{i,d} (\btheta_{0,i}) }{\partial \btheta_i }   \right \| _{\max}   + C s_{\beta} s_p \tau_m ^{1-\delta}.
 \end{equation*}
 Since $\bw_d$'s are bounded random variables and geometrically $\alpha$-mixing, 
 $ \( w_{i,d} - g_{i,d} (\btheta_{0i}) \) \times  \frac{  g_{i,d} (\btheta_{0,i}) }{\partial \btheta_i } = \epsilon_{i,d} \frac{  g_{i,d} (\btheta_{0,i}) }{\partial \btheta_i } $ is a martingale difference that is a  bounded random variable and geometrically $\alpha$-mixing satisfying Assumption \ref{assumption3}(c).
 Thus,  by Theorem 2 in \citet{merlevede2009bernstein}, we have, with probability at least $1- (pq)^{c}$, 
 \begin{equation*}
 	\left \|  \frac{2 }{n} \sum_{d=1}^n \( w_{i,d} - g_{i,d} (\btheta_{0i}) \)  \frac{  g_{i,d} (\btheta_{0,i}) }{\partial \btheta_i }   \right \| _{\max} \leq C  n ^{-1/2} \sqrt{ \log pq  }. 
 \end{equation*}
 Therefore, with Assumption \ref{assumption3}(e), by Proposition 1 in \citet{fan2019adaptive}, we can show \eqref{thm2-result}.
 \endpf
  
\subsection{Proof of Theorem \ref{thm-3}}

  \textbf{Proof of Theorem \ref{thm-3}.}
  Consider \eqref{thm3-result1}.
  We have 
  \begin{eqnarray*}
 |\hat{g}_{i, n+1} ( \hat{\btheta}_i)  - g_{i, n+1}( \btheta_{0,i}) | &\leq&  \| \hat{\btheta}_i - \btheta_{0i} \| _1  \max_{j \leq n} \| \hat{\bw}_{j} - \bw_{j} \|_{\max}  \cr
 	&& +  \| \btheta_{0i} \| _1  \max_{j \leq n} \| \hat{\bw}_{j} - \bw_{j} \|_{\max} +  \| \hat{\btheta}_i - \btheta_{0i} \| _1 \max_{j \leq n} \|  \bw_{j} \|_{\max} \cr
 	&\leq&   C   \frac{s_{\beta}  \lambda_n }{\kappa}  s_p \tau_m ^{1-\delta} + C s_{\beta}   s_p \tau_m ^{1-\delta}+  C   \frac{s_{\beta}  \lambda_n }{\kappa} \cr
 	&\leq& 	 C   \frac{s_{\beta}  \lambda_n }{\kappa} , 
  \end{eqnarray*}
  where the second inequality is due to Theorem \ref{thm-2} and Assumption \ref{assumption3}(d).
  
  Consider \eqref{thm3-result2}.
  We have 
  \begin{eqnarray*}
  && \left \|  \hat{ \bar{\bw}}_{n+1} - \frac{ \bg_{n+1} (\btheta_0)}{\b1^{\top} \bg_{n+1} (\btheta_0) } \right \|_1 \cr
  &&\leq  \left \|   \frac{ \hat{\bg}_{n+1} (\hat{\btheta}) -\bg_{n+1} (\btheta_0)}{\b1^{\top} \bg_{n+1} (\btheta_0) }  \right \|_1 + \left \|  \hat{\bg}_{n+1} (\hat{\btheta})  \right \| _1 \left \| \frac{  \b1^{\top} \bg_{n+1} (\btheta_0) -  \b1^{\top} \hat{\bg}_{n+1} (\hat{\btheta}) }{  \b1^{\top} \hat{\bg}_{n+1} (\hat{\btheta})  \b1^{\top} \bg_{n+1} (\btheta_0) } \right \|_1  \cr
  	&&\leq   C   \frac{s_{\beta}  \lambda_n }{\kappa} ,
  \end{eqnarray*}
  where the last inequality is due to \eqref{thm3-eq1} and \eqref{thm3-eq2} below.   
 We have
 \begin{eqnarray}\label{thm3-eq1}
 	\left \|   \frac{ \hat{\bg}_{n+1} (\hat{\btheta}) -\bg_{n+1} (\btheta_0)}{\b1^{\top} \bg_{n+1} (\btheta_0) }  \right \|_1 &\leq&  C p^{-1} 	\left \|    \hat{\bg}_{n+1} (\hat{\btheta}) -\bg_{n+1} (\btheta_0) \right \|_1 \cr
 		&\leq&  C   \frac{s_{\beta}  \lambda_n }{\kappa},
 \end{eqnarray} 
  where the first and second inequalities are due to Assumption \ref{assumption1}(a) and \eqref{thm3-result1}, respectively.  
We have
\begin{eqnarray}\label{thm3-eq2}
&& \left | \frac{  \b1^{\top} \bg_{n+1} (\btheta_0) -  \b1^{\top} \hat{\bg}_{n+1} (\hat{\btheta}) }{  \b1^{\top} \hat{\bg}_{n+1} (\hat{\btheta})  \b1^{\top} \bg_{n+1} (\btheta_0) } \right |  \cr
&& \leq   \frac{ |   \b1^{\top} \bg_{n+1} (\btheta_0) -  \b1^{\top} \hat{\bg}_{n+1} (\hat{\btheta}) |  }{  \[ \left | \b1^{\top} \{ \hat{\bg}_{n+1} (\hat{\btheta}) - \bg_{n+1} (\btheta_0) \} \right | +    \b1^{\top} \bg_{n+1} (\btheta_0)   \]   \b1^{\top}  \bg_{n+1} (\btheta_0) | }    \cr
&& \leq C p^{-1} \frac{s_{\beta}  \lambda_n }{\kappa},
\end{eqnarray}  
where the last inequality is due to Assumption \ref{assumption1}(a) and the fact that we have
\begin{eqnarray*}
	|  \b1^{\top} \bg_{n+1} (\btheta_0) -  \b1^{\top} \hat{\bg}_{n+1} (\hat{\btheta}) | &\leq& \| \bg_{n+1} (\btheta_0) -    \hat{\bg}_{n+1} (\hat{\btheta})  \|_1 \cr
		&\leq&    C   p \frac{s_{\beta}  \lambda_n }{\kappa},
\end{eqnarray*}  
where the last inequality is due to  \eqref{thm3-result1}.
Therefore, the statement is shown by Propositon \ref{prop1}.
  \endpf

\end{document}